\documentclass[12pt,aps,prd,onecolumn,nofootinbib]{revtex4-2}
\usepackage{amsmath,amssymb,bm}
\usepackage{graphicx}
\usepackage{hyperref}
\usepackage{physics}
\usepackage{xcolor}
\usepackage{times}
\usepackage{aas_macros}

\begin{document}

\title{The Scoured Spike: Suppression of Indirect Dark Matter Signals\\ by a Hidden Companion}

\author{Jaden Lopez}
\affiliation{Department of Physics, University of California, Santa Cruz, CA 95064, USA}
\author{Stefano Profumo}
\affiliation{Department of Physics, University of California, Santa Cruz, CA 95064, USA}
\affiliation{Santa Cruz Institute for Particle Physics, University of California, Santa Cruz, CA 95064, USA}


\begin{abstract}

A massive ``dark companion''---such as an intermediate-mass black hole or other compact dark object---orbiting the supermassive black hole at the Galactic Center can dynamically reshape the surrounding dark-matter spike. Through gravitational heating and angular-momentum exchange, the companion excavates a ``scoured'' region that lowers the inner density and suppresses the expected annihilation signal. We quantify this effect by computing the suppression of the dark-matter annihilation $J$-factor induced by such a companion, combining an analytic scouring-radius model with full numerical integrations of the modified density profile. We scan the parameter space of companion mass, orbital separation, system age, and spike slope, explicitly including the interplay with the annihilation plateau. \textcolor{black}{For canonical Gondolo--Silk spikes with $\gamma_{\rm sp} \gtrsim 2$, the scouring radius grows only weakly with injected energy because the binding energy is concentrated at small radii; however, once scouring occurs, the $J$-factor suppression can be substantial, for sufficiently massive or long-lived companions. In the shallower-spike regime with $\gamma_{\rm sp} \lesssim 2$, the same companion drives $r_{\rm scour}$ to much larger multiples of $R_{\rm core}$, and even a modest ${\sim}10^{4}\,M_\odot$ companion on a $\mathcal{O}(100)\,{\rm AU}$ orbit and $\mathcal{O}({\rm Gyr})$ age can suppress the annihilation flux by one to two orders of magnitude. Whether a steep spike is significantly suppressed therefore depends on the companion's mass and lifetime, not merely on the sign of $(2-\gamma_{\rm sp})$.} The numerical results are accurately captured (typically at the $\lesssim 10\%$ level) by a simple fitting formula in terms of a dimensionless scouring parameter that measures the ratio between the scoured region and the annihilation core. Our findings demonstrate that neglecting a dark companion can lead to substantial overestimates of the Galactic Center $J$-factor, with direct consequences for interpreting gamma-ray, neutrino, and antimatter searches for annihilating dark matter.

\end{abstract}

\maketitle

\section{Introduction}
\label{sec:intro}

The search for dark matter (DM) through its annihilation or decay products is one of the central goals of astroparticle physics. Indirect detection experiments aim to identify gamma rays, neutrinos, or cosmic-ray antimatter produced by DM interactions in astrophysical environments where the DM density is high. The expected flux from annihilating dark matter is proportional to the so-called \emph{$J$ factor}, the integral of the squared DM density $\rho_{\chi}$ along the line of sight. The $J$ factor encodes the entire astrophysical uncertainty of the indirect signal, independent of particle physics. Its precise value can vary by many orders of magnitude depending on the assumed distribution of DM in the inner Galaxy.

The Galactic Center (GC) has long been recognized as the most promising target for indirect searches due to its proximity and large expected DM density. Early analyses of \emph{Fermi} Large Area Telescope data revealed an extended excess of GeV gamma rays in the GC region~\cite{Goodenough:2009gk,Daylan:2014rsa,Calore:2015oya,Ajello:2016sxc,DiMauro:2021raz}. Although the spectral shape and morphology of this ``Galactic Center excess'' (GCE) are consistent with possible DM annihilation, conventional astrophysical interpretations involving millisecond pulsars or cosmic-ray outbursts remain viable~\cite{Bartels:2018qgr,Leane:2019xiy}. The viability of the DM interpretation depends sensitively on the adopted $J$ factor, which in turn depends on the inner slope and normalization of the Milky Way’s DM density profile.

In standard cold DM cosmology, the halo profile is often modeled as a Navarro--Frenk--White (NFW) or Einasto profile~\cite{Navarro:1996gj,Navarro:2008kc}. High-resolution $N$-body simulations have established a ``cuspy'' inner slope $\rho_{\chi}\propto r^{-\gamma}$ with $\gamma\simeq 1$ for collisionless DM~\cite{Springel:2008cc,Diemand:2008in}. However, the situation near the Galactic Center remains highly uncertain. Baryonic contraction can steepen the profile~\cite{Blumenthal:1985qy,Gnedin:2004cx}, while stellar feedback and dynamical heating can flatten it~\cite{Pontzen:2011ty,Governato:2012fa}. Estimates of the GC $J$ factor thus span several orders of magnitude~\cite{Hooper:2013rwa,Abazajian:2020tww,Karwin:2021jzp}, limiting the ability of indirect searches to constrain or infer DM particle properties.

Additional complexity arises from the presence of the central supermassive black hole, Sgr~A$^*$, which can induce a dense DM ``spike'' through adiabatic contraction~\cite{Gondolo:1999ef}. Such a spike can enhance the $J$ factor dramatically, boosting the annihilation flux by several orders of magnitude. Yet this configuration is dynamically fragile: stellar scattering, mergers, or the presence of secondary massive objects can erode or ``scour'' the spike~\cite{Merritt:2002vj,Gnedin:2003rj,Vasiliev:2007zx,Fields:2014pia}. The degree to which the Milky Way’s inner halo has retained or lost its spike remains unknown, but it has critical implications for the interpretation of current and future indirect-detection data.

In this work, we investigate how the presence of a \emph{dark companion}---a secondary compact object such as an intermediate-mass black hole or dark-matter–dominated clump---can alter the DM spike around Sgr~A$^*$. We show that such a companion can ``scour'' the inner density profile, reducing the $J$ factor and thus suppressing the predicted annihilation flux from the Galactic Center. Quantifying this suppression is essential for evaluating the robustness of indirect-detection constraints and for understanding whether the absence of an annihilation signal necessarily disfavours thermal dark matter.

\section{Indirect Detection and the Role of the \texorpdfstring{$J$}{J} Factor}
\label{sec:Jfactor}

Indirect searches for dark matter aim to detect the secondary particles produced by annihilation or decay of dark matter in regions of high density. For annihilating dark matter, the differential flux of gamma rays observed from a direction $\psi$ in the sky is given by
\begin{equation}
\frac{d\Phi_\gamma}{dE}(\psi) = 
\frac{\langle\sigma v\rangle}{8\pi m_\chi^2}\,
\frac{dN_\gamma}{dE}\,
J(\psi),
\end{equation}
where $\langle\sigma v\rangle$ is the velocity-averaged annihilation cross section, $m_\chi$ the particle mass, and $dN_\gamma/dE$ the photon yield per annihilation. All astrophysical uncertainties are contained in the \emph{$J$ factor},
\begin{equation}
J(\psi) = \int_{\text{l.o.s.}} \rho_\chi^2[r(s,\psi)]\, ds .
\end{equation}
For decaying dark matter the flux depends linearly on $\rho_\chi$ rather than $\rho_\chi^2$, but the Galactic Center remains the dominant source for both cases.

\subsection{Importance for gamma rays, antimatter, and neutrinos}

The $J$ factor directly controls the normalization of the expected signal across all indirect probes. For gamma rays, it determines the intensity of the prompt annihilation flux as well as the secondary emission from inverse-Compton scattering and bremsstrahlung~\cite{Cirelli:2010xx,Bringmann:2012ez,Slatyer:2015jla}. For cosmic-ray antimatter (positrons, antiprotons) the local annihilation rate depends on the average of $\rho_\chi^2$ convolved with propagation effects, but uncertainties in the Galactic distribution still propagate into the inferred cross-section limits~\cite{Donato:2008jk,DiMauro:2014pqa,Boudaud:2014qra}. For neutrinos, line-of-sight integrals analogous to the $J$ factor enter the expected flux in detectors such as IceCube and KM3NeT~\cite{Abbasi:2021xzx,Adrian-Martinez:2016fei}.

In all cases, the scaling with $\rho_\chi^2$ implies that even modest variations in the inner density profile of the Milky Way can alter the expected flux by several orders of magnitude. Thus, uncertainties in the central DM distribution translate directly into uncertainties in the inferred annihilation cross section or lifetime.

\subsection{Standard estimates, uncertainties, and the innermost halo profile}
\label{subsec:stdJ}

For an NFW profile normalized to the local density $\rho_\odot\simeq
0.4~\mathrm{GeV\,cm^{-3}}$ and scale radius $r_s\simeq20~\mathrm{kpc}$, the
canonical $J$ factor, averaged on a $1^\circ$ region around the Galactic Center, is
\[
J_{\rm NFW}\sim 10^{22-23}~\mathrm{GeV^2\,cm^{-5}}
\]
\cite{Hooper:2013rwa,Calore:2015oya}.  
Steeper cusps or central ``spikes’’ can increase this value by several orders of
magnitude~\cite{Gondolo:1999ef,Fields:2014pia}, whereas cored or dynamically
heated profiles can suppress it significantly
\cite{Pontzen:2011ty,Governato:2012fa,Chan:2015tna}.  
High-resolution $N$-body simulations
(\textsc{Via Lactea}, \textsc{Aquarius}, \textsc{Illustris})
consistently predict cuspy profiles with $\rho_\chi\propto r^{-\gamma}$ and
$\gamma\simeq 1$ in the absence of baryons
\cite{Navarro:1996gj,Navarro:2008kc,Springel:2008cc,Diemand:2008in},
but baryonic physics substantially reshapes the innermost density structure.

\subsubsection*{Adiabatic contraction and baryonic effects}

The slow accumulation of baryons in the central Galaxy---via the bulge, bar,
and nuclear star cluster---steepens the dark-matter cusp through
\emph{adiabatic contraction}.  
When the gravitational potential deepens on timescales long compared to orbital
periods, adiabatic invariants are approximately conserved and DM orbits contract
\cite{Blumenthal:1985qy,Gnedin:2004cx}.  
Semi-analytic models and hydrodynamic simulations often find effective inner
slopes $\gamma\gtrsim1$ once baryons are included.

However, stellar and AGN feedback, clumpy accretion, black-hole scattering, and
relaxation processes can flatten or even erase central cusps
\cite{Pontzen:2011ty,Governato:2012fa,Chan:2015tna,Merritt:2004xa}.  
As a result, realistic Milky-Way models span a broad range of inner slopes,
$\gamma\sim0.6$--$1.4$, introducing order-of-magnitude variations in the $J$
factor even before accounting for the central black hole.

\subsubsection*{The Gondolo--Silk spike: adiabatic growth of Sgr~A$^*$}

If Sgr~A$^*$ formed or grew \emph{adiabatically} at the center of a pre-existing
cuspy halo with slope $\gamma$, the phase-space density rearranges into a much
steeper ``spike’’ within a characteristic radius $R_{\rm sp}\sim\mathcal{O}(0.1)\,r_h$,
where $r_h$ is the black-hole influence radius \cite{Gondolo:1999ef}.  
The resulting spike follows the analytic power law
\begin{equation}
\rho_{\rm sp}(r) \propto r^{-\gamma_{\rm sp}}, 
\qquad
\gamma_{\rm sp} = \frac{9 - 2\gamma}{4 - \gamma},
\label{eq:GS_gamma}
\end{equation}
which gives $\gamma_{\rm sp}\simeq2.25$ for an NFW progenitor with
$\gamma\simeq1$.  

Because the annihilation luminosity scales as $\rho^2$, a steep spike can boost
the $J$ factor by several orders of magnitude compared with an NFW or Einasto
halo normalized to the same local density
\cite{Gondolo:1999ef,Fields:2014pia,Shapiro:2016ypb}.  
In such scenarios the Milky Way would host an exceptionally bright annihilation
region within the central $\sim 0.1$~pc.

\subsubsection*{Annihilation-limited density and the core radius}

At sufficiently small radii, annihilations deplete the spike and enforce a
maximum density determined by the system's age $t_{\rm age}$:
\begin{equation}
\rho_{\rm core} \simeq 
\frac{m_\chi}{\langle\sigma v\rangle\,t_{\rm age}}.
\label{eq:rhocore}
\end{equation}
The \emph{annihilation core radius} $R_{\rm core}$ is given implicitly by
\begin{equation}
\rho_{\rm sp}(R_{\rm core}) = \rho_{\rm core} .
\label{eq:Rcore}
\end{equation}
The complete unperturbed profile is conveniently expressed in piecewise form as
\begin{equation}
\rho_\chi(r) \simeq
\begin{cases}
\rho_{\rm core}, & r \lesssim R_{\rm core}, \\[4pt]
\rho_{\rm sp}(r)\propto r^{-\gamma_{\rm sp}}, & 
R_{\rm core} \lesssim r \lesssim R_{\rm sp}, \\[4pt]
\rho_{\rm halo}(r), & r \gtrsim R_{\rm sp},
\end{cases}
\label{eq:piecewise_spike_profile}
\end{equation}
where $\rho_{\rm halo}(r)$ matches onto the kiloparsec-scale NFW/Einasto halo.

The $J$ factor from a steep spike is dominated by radii near $R_{\rm core}$,
so even modest changes in $R_{\rm core}$ or $\gamma_{\rm sp}$ can result in
order-of-magnitude differences in the predicted annihilation flux.

\subsubsection*{Dynamical fragility of the spike}

The canonical Gondolo--Silk spike is dynamically fragile.  
Scattering with stars and stellar-mass black holes in the nuclear cluster
drives energy relaxation that gradually softens the spike over $\sim$Gyr
timescales \cite{Merritt:2002vj,Gnedin:2003rj,Vasiliev:2007zx}.  
Past mergers and the inspiral of intermediate-mass black holes can partially or
fully destroy the spike, producing shallower cusps or central cores
\cite{Merritt:2006mt,Gualandris:2010,Thomas:2014,Bortolas:2016}.  
Thus the effective slope $\gamma_{\rm sp}$ in realistic nuclei may lie anywhere
between $\sim 2.25$ (adiabatic) and $\sim1.5$ (relaxed), spanning several orders
of magnitude in the corresponding $J$ factor.

\subsubsection*{Connection to the present work}

Indirect-detection analyses including the
\textit{Fermi}-LAT Galactic Center excess~\cite{Daylan:2014rsa,Calore:2015oya,DiMauro:2021raz}
and H.E.S.S./CTA~\cite{Abdallah:2016ygi,CTAConsortium:2018tzg} almost always
assume an unperturbed NFW cusp or a mildly contracted halo.  
Yet the central few parsecs of the Milky Way are dynamically complex: stellar
heating, relaxation, and possible intermediate-mass companions can reshape the
innermost profile dramatically.

In this work we demonstrate that a dark companion can further enlarge the
low-density region---effectively increasing $R_{\rm core}$---and suppress the
$J$ factor by orders of magnitude, particularly when the underlying spike is
already shallower than the adiabatic $\gamma_{\rm sp}\simeq2.25$ value.  
The analytical expressions above therefore provide the natural baseline against
which the dynamical effects in subsequent sections should be interpreted.

\section{Dark Companion Parameter Space and Observational Constraints}
\label{sec:constraints}

The possibility that Sgr~A$^*$ is accompanied by a secondary compact object---a ``dark companion''---has attracted increasing attention as both a natural outcome of hierarchical galaxy formation and a potentially transformative element in the interpretation of indirect dark matter searches.  In this section we review the physical and observational constraints on such a companion, summarize the viable parameter space in terms of its mass and orbital separation, and discuss the implications for dynamical stability, gravitational-wave observability, and astrophysical signatures.

\subsection{Motivation and formation channels}

The hierarchical assembly of galaxies virtually guarantees that the Milky Way has undergone multiple mergers in the past few Gyr, implying that the formation of a binary black hole (BBH) system in its nucleus is a natural possibility \cite{Kormendy:1995yz,Lotz:2011av,Unavane:1996}.  Dynamical friction acting on the smaller black hole during the merger of its host galaxy drives it toward the galactic center, where it can form a bound pair with Sgr~A$^*$.  Depending on the mass ratio, initial separation, and interaction with the nuclear stellar cluster and surrounding gas, the binary may stall at a parsec or sub-parsec separation, or proceed to coalescence via gravitational-wave emission \cite{Merritt:2002vj,Roedig:2011}.  

{While the coalescence timescale for an equal-mass supermassive binary can
be shorter than a Hubble time under favorable conditions (e.g.\ efficient
gas dissipation or triaxial stellar potentials), such systems may also
stall at parsec or sub-parsec separations due to the well-known
final-parsec problem. Importantly, our analysis does not rely on rapid
coalescence in this regime. By contrast, intermediate-mass companions
($10^2$--$10^5\,M_\odot$) can remain bound for gigayears at separations of
$\mathcal{O}(10$--$10^3)\,$AU, producing potentially measurable dynamical
signatures without contradicting stellar orbital data\footnote{The long-term evolution of supermassive black-hole binaries
depends sensitively on the stellar and gaseous environment; stalling at
parsec scales remains a well-known possibility in collisionless nuclei.}.
}

\subsection{Dynamical and astrometric constraints}

The orbits of the S-stars around Sgr~A$^*$ provide some of the tightest constraints on any hypothetical companion.  Detailed analyses of S2’s precession and radial velocity curve \cite{Ghez:1998,Gravity:2020} limit perturbations to the gravitational potential at the $\sim1\%$ level within $\sim 1000$~AU.  Building upon these data, Naoz \textit{et al.}~\cite{Naoz:2019} mapped the region of stability in the $(m_2, a_2)$ plane for a secondary black hole of mass $m_2$ and semi-major axis $a_2$, showing that companions with masses between $\sim10^2$ and $10^5\,M_\odot$ at separations between $10$ and $10^4\,\mathrm{AU}$ remain dynamically allowed.  The excluded regions correspond either to systems that would strongly perturb the observed S-star orbits or to configurations that would be tidally disrupted or have merged on short timescales.

Additional constraints arise from astrometric monitoring of Sgr~A$^*$’s own motion relative to the radio reference frame, which limits any reflex motion due to a companion.  VLBA and EHT data constrain the amplitude of such motion to below $\sim 1\,\mathrm{km\,s^{-1}}$, excluding companions more massive than $\sim10^5\,M_\odot$ at $\lesssim 10^3\,\mathrm{AU}$ separations \cite{Reid:2004,EHT:2022}.  However, this still leaves open a substantial window of parameter space where a dark companion could exist undetected.

\begin{figure}[t]
\centering
\fbox{\includegraphics[width=0.9\textwidth]{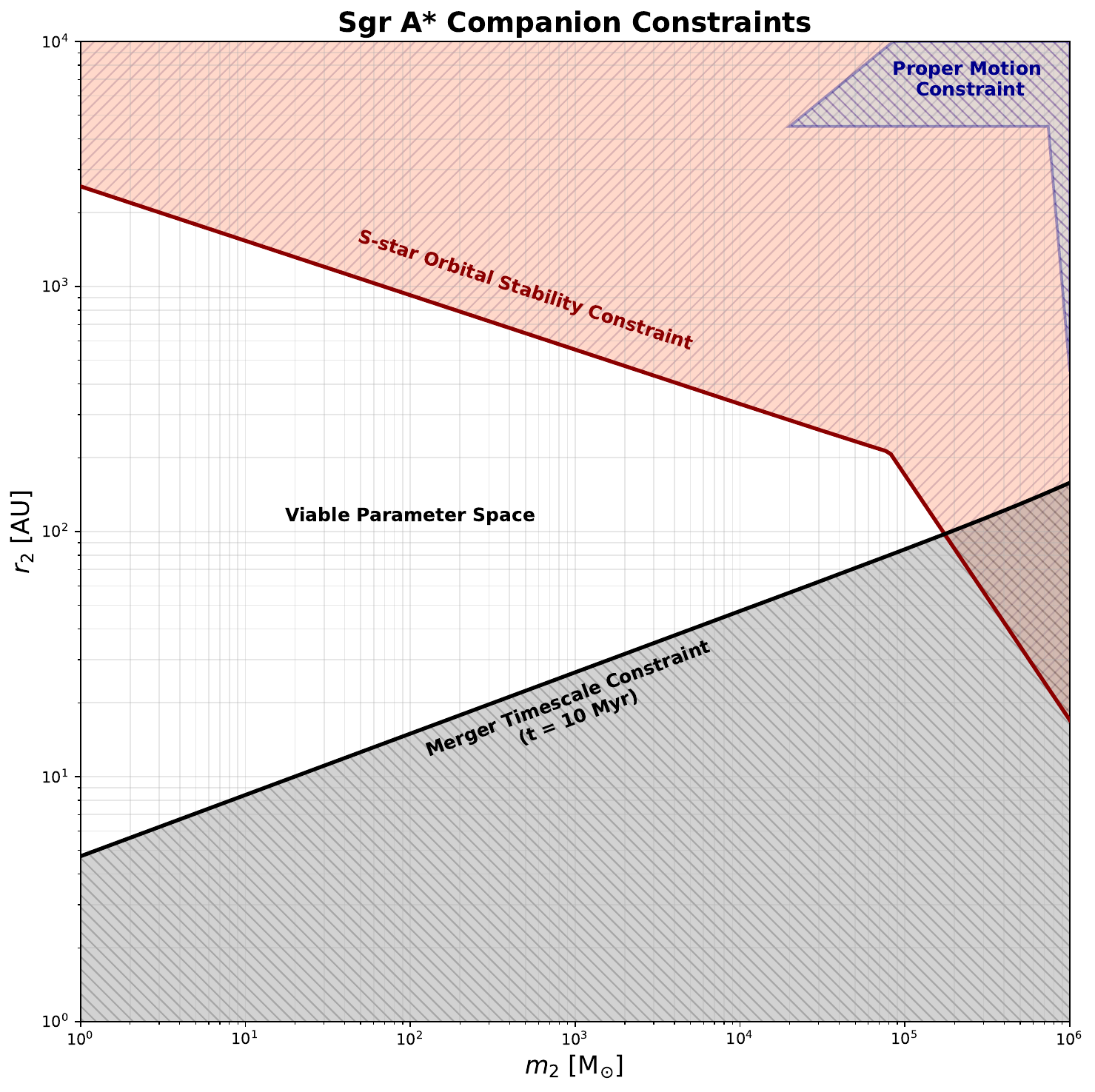}}
\caption{Observational constraints on a hypothetical dark companion to Sgr A*. Shaded regions indicate exclusions from: (red) S-star orbital stability requiring $<1$\% perturbations to the gravitational potential \cite{Naoz:2019}; (blue) proper motion limits from VLBA/EHT astrometry \cite{Reid:2004,EHT:2022} constraining reflex motion $<1 km$ $s^{-1}$; (gray) gravitational-wave merger timescale $<10$ Myr, incompatible with long-lived scouring \cite{Naoz:2019}. The unshaded central region corresponds to the viable parameter space explored in this work.}
\label{fig:constraints}
\end{figure}

\subsection{Gravitational-wave and accretion constraints}

Companions in the $10^3$–$10^5\,M_\odot$ range with separations of tens to thousands of astronomical units would evolve on timescales of $10^7$–$10^{10}$~yr, implying quasi-stationary configurations from a galactic perspective.  Their gravitational-wave strain lies below the sensitivity of current pulsar timing arrays but within reach of the planned \textit{LISA} mission \cite{LISA:2017}.  Detection of an IMRI signal from the Galactic Center would not only confirm the existence of a secondary object but would also directly constrain the total binding energy available to reshape the dark matter spike.

Accretion signatures are expected to be minimal if the companion is nonluminous or embedded within the radiatively inefficient accretion flow of Sgr~A$^*$; in this case, the companion would be effectively ``dark'' in electromagnetic observations, consistent with current infrared and X-ray limits \cite{Gillessen:2017}.

\subsection{Parameter space summary}

Combining all current constraints, the viable parameter space for a dark companion is bounded approximately by
\[
10^2\,M_\odot \lesssim m_2 \lesssim 10^5\,M_\odot, \qquad
10\,\mathrm{AU} \lesssim a_2 \lesssim 10^4\,\mathrm{AU},
\]
with the precise boundaries depending on orbital inclination, eccentricity, and the detailed modeling of the nuclear stellar cluster.  This range encompasses the region where dynamical friction and gravitational-wave emission are comparable over a gigayear, leading to significant energy transfer to the surrounding dark matter.  As we show in subsequent sections, such companions are capable of scouring the central dark matter spike over radii comparable to the canonical annihilation core, thereby suppressing the effective $J$ factor by orders of magnitude.

\section{Analytical Framework: The Scouring Radius Approximation}
\label{sec:scour}

The presence of a secondary compact object in the Galactic Centre fundamentally alters the equilibrium configuration of the dark matter (DM) density spike around the primary black hole.  The orbital motion and dynamical friction associated with this \emph{dark companion} deposit kinetic energy into the surrounding medium, leading to a partial evacuation, or ``scouring,'' of the inner density cusp.  To describe this effect quantitatively, we adopt the semi‐analytical framework developed in Refs.~\cite{Merritt:2002vj,Gnedin:2003rj,Vasiliev:2007zx,Fields:2014pia,Amaro-Seoane:2004nt,Profumo:2025prep}, which encapsulates the impact of the companion through a characteristic \emph{scouring radius} $r_{\rm scour}$.  This approach provides an intuitive and computationally tractable description of the reduction of the astrophysical $J$ factor discussed in Section~\ref{sec:Jfactor}.

\subsection{Energy balance and the definition of the scouring radius}

The scouring radius is defined by the condition that the total orbital energy injected by the secondary object over its lifetime equals the binding energy of the DM contained within $r_{\rm scour}$.  Specifically,
\begin{equation}
\Delta E_{\rm DF}(r_2) \simeq |U_{\rm bind}(r_{\rm scour})| ,
\end{equation}
where $\Delta E_{\rm DF}$ is the cumulative energy transferred to the spike through dynamical friction as the companion of mass $m_2$ inspirals from its initial orbital radius $r_2$, and $U_{\rm bind}$ is the gravitational binding energy of the spike interior to $r_{\rm scour}$ with respect to the primary black hole of mass $m_1$.


{
The orbital energy dissipated by dynamical friction is given approximately by \cite{BinneyTremaine,Antonini:2012ad}
\begin{equation}\label{eq8}
\Delta E_{\rm DF} \approx 4\pi (G m_2)^2 \int_0^{t_{\rm bin}}
\frac{\rho_{\chi}(r_2(t))}{v(t)} \ln\Lambda \, dt ,
\end{equation}
where $\rho_{\chi}(r)$ is the ambient DM density, $v \simeq \sqrt{G m_1/r_2}$ is the orbital
velocity, and $t_{\rm bin}$ is the binary lifetime. The Coulomb logarithm $\ln\Lambda$ depends on
the ratio of maximum to minimum impact parameters and is therefore uncertain at the
order-unity level in galactic nuclei. We adopt a fiducial value $\ln\Lambda = 10$, which lies within
the range reported in the literature; varying $\ln\Lambda$ within this range changes the inferred
scouring radius only logarithmically and does not affect the qualitative conclusions of this work.
}

{
When we treat the companion as long-lived on a quasi-stationary orbit at radius $r_2$ (so that $\rho_{\chi}$ and $v$ are evaluated at $r_2$ and taken approximately constant over $t_{\rm bin}$), Eq.~(\ref{eq8}) reduces to the controlled estimate
\[
\Delta E_{\rm DF}\;\approx\;4\pi (Gm_2)^2\,\rho_{\chi}(r_2)\,\frac{\ln\Lambda}{v(r_2)}\,t_{\rm bin}.
\]
 Conversely, if one wishes to model orbital decay $r_2(t)$ explicitly, the evolution must be obtained by solving the dynamical-friction equation of motion for $r_2(t)$ and then evaluating the time integral in Eq.~(\ref{eq8}).

}

Balancing this with the gravitational binding energy of the spike between $r_{\rm core}$ (the annihilation core radius) and $r_{\rm scour}$ yields
\begin{equation}
|U_{\rm bind}(r_{\rm scour})|
\simeq 4\pi G m_1 \rho_R R_{\rm sp}^{\gamma_{\rm sp}}
\frac{r_{\rm core}^{2-\gamma_{\rm sp}} - r_{\rm scour}^{2-\gamma_{\rm sp}}}{2-\gamma_{\rm sp}},
\end{equation}
where $\rho_R$ and $R_{\rm sp}$ are the characteristic density and radius of the unperturbed spike {($\rho_R$ is the DM density at $R_{\rm sp}$)}, and $\gamma_{\rm sp}$ is its logarithmic slope. {Note that in the simplified expression above, the Coulomb logarithm can be absorbed into an overall
order-unity normalization and is retained explicitly in all numerical evaluations.
} Solving for $r_{\rm scour}$ gives
\begin{equation}
r_{\rm scour} \simeq 
\left[
r_{\rm core}^{2-\gamma_{\rm sp}}
+ \frac{(\gamma_{\rm sp}-2)\,\Delta E_{\rm DF}}{4\pi G m_1 \rho_R R_{\rm sp}^{\gamma_{\rm sp}}}
\right]^{1/(2-\gamma_{\rm sp})},
\label{eq:rscour}
\end{equation}
which reduces to the Gondolo–Silk spike ($r_{\rm scour}\to r_{\rm core}$) when $\Delta E_{\rm DF}\to 0$.  

{
Equation~(\ref{eq:rscour}) follows from equating the dynamical-friction energy deposited by the
companion, $\Delta E_{\rm DF}$, to the change in gravitational binding energy of the spike between
$r_{\rm core}$ and $r_{\rm scour}$. For a spike profile
$\rho_{\rm sp}(r)=\rho_R (r/R_{\rm sp})^{-\gamma_{\rm sp}}$ with $\gamma_{\rm sp}>1$, the change in
binding energy satisfies
\begin{equation}
\Delta E_{\rm bind}(r_{\rm scour}) \equiv
E_{\rm bind}(r_{\rm scour})-E_{\rm bind}(r_{\rm core})
\propto r_{\rm scour}^{\,2-\gamma_{\rm sp}}-r_{\rm core}^{\,2-\gamma_{\rm sp}},
\end{equation}
which is a monotonically increasing function of $r_{\rm scour}$ for $\gamma_{\rm sp}<2$ and remains
well defined for $\gamma_{\rm sp}>2$ in the sense of energy differences.

Since $\Delta E_{\rm DF}\ge 0$ by construction, energy balance requires
$\Delta E_{\rm bind}(r_{\rm scour})\ge 0$, which immediately implies
$r_{\rm scour}\ge r_{\rm core}$. Algebraically, solving the energy-balance equation therefore yields
two formal roots, but only the branch with
\begin{equation}\label{eq:sdef}
r_{\rm scour}\ge r_{\rm core}\qquad (s\equiv r_{\rm scour}/r_{\rm core}\ge 1)
\end{equation}
satisfies both energy conservation and the physical requirement that dynamical heating cannot
reduce the size of a pre-existing annihilation plateau. The alternative root corresponds to
$r_{\rm scour}<r_{\rm core}$ and would imply a negative injected energy; it is therefore unphysical
and is discarded.
}

\subsection{Modified density profile and annihilation suppression}\label{sec:Jmod}

The simplest phenomenological description of the scoured profile is
\begin{equation}
\rho_{\rm mod}(r) = \rho_{\rm sp}(r)\,\Theta(r - r_{\rm scour}),
\label{eq:rho_mod}
\end{equation}
where $\Theta$ is the Heaviside step function and $\rho_{\rm sp}(r)$ is the preexisting spike density, $\rho_{\rm sp}\propto r^{-\gamma_{\rm sp}}$.  
In practice, numerical N-body experiments and Fokker–Planck calculations suggest a smoother transition, with a gradual flattening below $r_{\rm scour}$ rather than a sharp cut \cite{Vasiliev:2007zx,Merritt:2010}.  However, the step-function model captures the integrated suppression of the $J$ factor with sufficient accuracy for our purposes, particularly in the regime $\gamma_{\rm sp}>2$ where the line-of-sight integral is dominated by the innermost region.

The corresponding annihilation $J$ factor for the scoured profile can be approximated analytically as\footnote{For $\gamma_{\rm sp} \ge 3/2$, the annihilation signal becomes increasingly dominated by the smallest surviving radius, and the suppression must be computed using the full density profile rather than a simple power-law truncation.}

\begin{equation}
\frac{J_{\rm mod}}{J_{\rm sp}}
\simeq
1 -
\left(
\frac{r_{\rm scour}}{R_{\rm sp}}
\right)^{3-2\gamma_{\rm sp}},
\qquad
\gamma_{\rm sp} < \frac{3}{2},
\label{eq:Jmod_valid}
\end{equation}

neglecting the core radius when $r_{\rm core}\ll r_{\rm scour}\ll R_{\rm sp}$.
Equation~\eqref{eq:Jmod_valid} follows directly from the step-function model in Eq.~\eqref{eq:rho_mod} and the scaling
$J\propto \int r^2 \rho_{\rm sp}^2(r)\,dr \propto \int r^{2-2\gamma_{\rm sp}}dr$ over the spike.
For $\gamma_{\rm sp}<3/2$ this integral is dominated by radii near the \emph{outer} edge of the spike,
$r\sim R_{\rm sp}$, and excising only the inner region produces a parametrically small suppression
when $r_{\rm scour}\ll R_{\rm sp}$, as made explicit by Eq.~(\ref{eq:Jmod_valid}).
The case $\gamma_{\rm sp}=3/2$ is marginal and yields a logarithmic dependence,
\begin{equation}
\left.\frac{J_{\rm mod}}{J_{\rm sp}}\right|_{\gamma_{\rm sp}=3/2}
\simeq
\frac{\ln\!\left(R_{\rm sp}/r_{\rm scour}\right)}{\ln\!\left(R_{\rm sp}/r_{\rm core}\right)}\,,
\label{eq:Jratio_gammasp_threehalves}
\end{equation}
so a non-negligible suppression is expected whenever $r_{\rm scour}>r_{\rm core}$.
For steeper spikes with $\gamma_{\rm sp}>3/2$, the integral becomes dominated by the smallest surviving
radius and Eq.~\eqref{eq:Jmod_valid} must not be extrapolated into this regime. In that case, when scouring enlarges the
effective inner cutoff from $r_{\rm core}$ to $r_{\rm scour}$, the suppression scales instead as
\begin{equation}
\left.\frac{J_{\rm mod}}{J_{\rm sp}}\right|_{\gamma_{\rm sp}>3/2}
\simeq
\left(\frac{r_{\rm scour}}{r_{\rm core}}\right)^{3-2\gamma_{\rm sp}}\,,
\label{eq:Jratio_gammasp_gt_threehalves}
\end{equation}
up to order-unity factors associated with the detailed shape of the inner profile.

\subsection{Parameter dependencies and scaling regimes}

Combining Eqs.~(\ref{eq:rscour}) and (\ref{eq:Jratio_gammasp_gt_threehalves})
\textcolor{black}{reveals several qualitative dependencies. In the regime $\gamma_{\rm sp} > 3/2$, where the $J$-factor integral is dominated by the smallest surviving radius, the suppression ratio satisfies $J_{\rm mod}/J_{\rm sp} \simeq (r_{\rm scour}/r_{\rm core})^{3-2\gamma_{\rm sp}}$ with an exponent that grows more negative as $\gamma_{\rm sp}$ increases. Steep spikes therefore exhibit large $J$-factor suppression \emph{per unit increase} in $r_{\rm scour}/r_{\rm core}$, even though $r_{\rm scour}$ itself grows more slowly with $\Delta E_{\rm DF}$ in the steep-spike regime. The net effect depends on the interplay of both factors}:

\begin{itemize}
\item Increasing the companion mass $m_2$ enhances $\Delta E_{\rm DF}\propto m_2^{3/2}$, pushing $r_{\rm scour}$ outward and suppressing $J$ through the scaling of Eq.~(\ref{eq:Jratio_gammasp_gt_threehalves}).  Companions with $m_2\gtrsim10^4\,M_\odot$ can therefore erase the spike entirely out to the annihilation plateau.
\item Larger orbital separations $r_2$ reduce the local DM density and hence $\Delta E_{\rm DF}$, yielding $r_{\rm scour}\simeq r_{\rm core}$ and negligible suppression.
\item Shorter binary lifetimes $t_{\rm bin}$ (e.g.\ due to gravitational radiation) decrease the available time for energy deposition and thus reduce $r_{\rm scour}$.
\item \textcolor{black}{Shallower spikes (smaller $\gamma_{\rm sp}$) allow a given $\Delta E_{\rm DF}$ to excavate a larger region because the central binding energy is lower; but once $r_{\rm scour}/r_{\rm core} \gg 1$, steeper spikes suffer comparably large or larger fractional $J$-factor losses because of the steeper scaling in Eq.~(\ref{eq:Jratio_gammasp_gt_threehalves}).}
\end{itemize}
These trends naturally explain the results of the numerical exploration presented in Section~\ref{sec:results}:  only compact, long-lived, and relatively massive companions can drive the system into the $s\gg1$ regime where the annihilation signal is substantially quenched.

\subsection{Validity and limitations}

The scouring-radius approximation is accurate as long as the companion’s orbit remains bound and circular and as long as the DM responds adiabatically to the energy injection.  For eccentric or time-variable orbits, the true energy transfer rate may be higher, producing a shallower but more extended depletion region.  Moreover, stellar encounters and gas drag can act in concert with the dark companion, reinforcing the overall suppression of the spike \cite{Gualandris:2010}.  A full kinetic treatment would require integrating the Fokker–Planck equation with self-consistent source terms for both stellar and black-hole heating, which is beyond the scope of this paper but under development in ongoing work \cite{Profumo:2025prep}.

Despite these caveats, the scouring-radius formalism provides a transparent link between the microphysics of dynamical heating and the macroscopic observables relevant for indirect detection.  In the next section we apply this framework to compute the suppression of the $J$ factor over the viable parameter space for dark companions identified in Section~\ref{sec:constraints}.

{We therefore explicitly exclude the regime $\gamma_{\rm sp} \ge 3/2$ from the application of the analytic $J$-factor scaling in Eq.~(\ref{eq:Jmod_valid}); steeper spikes require either a full numerical evaluation or a modified treatment of the inner cutoff.
}

\section{Dependence on the Dark Matter Spike and Core}
\label{sec:spikecore}

The scouring induced by a dark companion depends sensitively on the structure of
the \emph{pre-existing} dark-matter distribution around Sgr~A$^*$.  
The results of the previous section established that the companion injects energy
into the inner halo, inflating the effective low-density region and suppressing
the annihilation signal.  
Here we examine how this suppression scales with the parameters that determine
the unperturbed profile—specifically the spike slope $\gamma_{\rm sp}$, the
annihilation core size, and the age of the system.  
Since the physical origins of the spike (adiabatic contraction, black-hole
growth, annihilation plateau formation, and dynamical heating) were already
discussed in Sec.~\ref{subsec:stdJ}, our goal in this section is not to revisit
that physics, but to quantify how variations in these parameters control the
magnitude of the companion-induced suppression of the $J$ factor.

\subsection{The role of the spike slope and the annihilation core}
\label{subsec:spikerole}

In the scouring framework, the spike slope $\gamma_{\rm sp}$ enters solely
through the binding energy profile of the cusp, which governs how difficult it
is for the companion to excavate material from small radii.  
\textcolor{black}{Steeper spikes ($\gamma_{\rm sp}>2$) concentrate mass and binding energy near the
center, so a given $\Delta E_{\rm DF}$ produces only a modest increase in $r_{\rm scour}$---this is the sense in which they are ``hard to scour.''
However, because Eq.~(\ref{eq:Jratio_gammasp_gt_threehalves}) governs the $J$-factor for $\gamma_{\rm sp}>3/2$, even a moderate increase in $r_{\rm scour}/r_{\rm core}$ can yield substantial $J$-factor suppression; the exponent $(3-2\gamma_{\rm sp})$ becomes more negative as $\gamma_{\rm sp}$ increases.}
Conversely, shallower spikes ($\gamma_{\rm sp}<2$), whether produced by stellar
heating or past dynamical activity, have much lower central binding energy and
are therefore highly susceptible to scouring.

The second key ingredient is the annihilation plateau.  
Even in the absence of a companion, annihilations enforce a minimum density
$\rho_{\rm core}$ and an associated core radius $R_{\rm core}$ that depend on
$m_\chi$, $\langle\sigma v\rangle$, and the system age $t_{\rm age}$.  
The companion effectively increases this core radius to the scouring radius
$r_{\rm scour}$ derived in Sec.~\ref{sec:scour}, so the relevant quantity that
controls the suppression is the ratio $s$ defined in Eq.~(\ref{eq:sdef}.
Because the annihilation signal is dominated by radii near $R_{\rm core}$ in
steep spikes, even modest increases in $s$ can substantially reduce the total
$J$ factor.  
This simple scaling explains many qualitative features of the numerical results
in Sec.~\ref{sec:results}.

\subsection{Transition from resilient to fragile spikes}

Inspection of Eq.~(\ref{eq:rscour}) shows that the response of the profile to a
fixed amount of injected energy changes qualitatively at $\gamma_{\rm sp}=2$.
For $\gamma_{\rm sp}>2$, the binding energy diverges toward the center and
$r_{\rm scour}$ depends only weakly on the companion's parameters.  
\textcolor{black}{In this ``hard-to-scour'' regime, $r_{\rm scour}$ grows slowly with $\Delta E_{\rm DF}$, so a more massive or longer-lived companion is required to drive $r_{\rm scour}$ appreciably above $r_{\rm core}$.
However, this does \emph{not} imply $J_{\rm mod}/J_{\rm sp}\approx1$: once $r_{\rm scour}$ exceeds $r_{\rm core}$ by even a moderate factor, the steep scaling of Eq.~(\ref{eq:Jratio_gammasp_gt_threehalves}) can produce substantial $J$-factor suppression.
For $\gamma_{\rm sp}\simeq2.25$, the exponent $(3-2\gamma_{\rm sp})\approx-1.5$ means that $r_{\rm scour}/r_{\rm core}=10$ already suppresses $J_{\rm mod}/J_{\rm sp}$ by roughly two orders of magnitude.
The ``resilient'' character of steep spikes should therefore be understood as resilience against \emph{large scouring radii}, not as immunity from $J$-factor suppression.}

For $\gamma_{\rm sp}<2$, the binding energy decreases toward small radii and the
dependence of $r_{\rm scour}$ on $\Delta E_{\rm DF}$ becomes strongly
nonlinear.  
Small changes in the companion's mass, orbital separation, or lifetime translate
into large increases in $r_{\rm scour}$ and correspondingly large decreases in
the annihilation flux.  
This ``fragile'' regime is the one most sensitive to the presence of a dark
companion, and it naturally arises in any spike that has undergone substantial
stellar heating or past dynamical activity.

\subsection{Interplay with particle-physics parameters}

Although the companion does not couple directly to particle-physics quantities,
the annihilation plateau couples them indirectly to scouring.  
Larger $m_\chi$, smaller $\langle\sigma v\rangle$, or longer system age all
raise $\rho_{\rm core}$ and shrink $R_{\rm core}$, effectively increasing the
degree of concentration of the spike.  
A more concentrated spike requires more energy to excavate and is therefore less
susceptible to scouring.  
This shows that astrophysical and particle-physics uncertainties are entangled:
the same system may exhibit strong or negligible suppression depending on the
assumed microphysical parameters.

The next section quantifies these trends through explicit numerical evaluation
of the line-of-sight integral for the scoured density profile across the
allowed companion parameter space.

\subsection{Transition to heated and shallower spikes}

Real galactic nuclei are not perfectly adiabatic.  Encounters with stars, transient binaries, and molecular clouds transfer energy to the DM, gradually softening the spike \cite{Merritt:2004xa,Amaro-Seoane:2004nt,Vasiliev:2007zx}.  
N-body and Fokker–Planck studies show that over $\sim10$ Gyr, a spike with $\gamma_{\rm sp}\simeq2.3$ can evolve to $\gamma_{\rm sp}\lesssim1.8$ or even approach an isothermal core if relaxation is efficient \cite{Vasiliev:2007zx,Gnedin:2003rj}.  
This transition has profound implications for the effect of a dark companion: in the shallow-spike regime, the gravitational binding energy is lower and the same dynamical friction energy can excavate a much larger region.

Equation~(\ref{eq:rscour}) illustrates this behavior explicitly.  
When $\gamma_{\rm sp}>2$, the binding energy diverges toward small radii, so the scouring radius changes slowly with $\Delta E_{\rm DF}$.  
For $\gamma_{\rm sp}<2$, by contrast, the exponent $(2-\gamma_{\rm sp})^{-1}$ becomes positive, and small changes in $\Delta E_{\rm DF}$ translate into large increases in $r_{\rm scour}$.  
Consequently, the ratio $s$, defined in Eq.~(\ref{eq:sdef}, rises steeply as $\gamma_{\rm sp}$ drops below~2, leading to an exponential-like suppression of the annihilation flux.

\subsection{Interplay between annihilation and dynamical scouring}

The competition between the annihilation plateau and the dynamical scouring determines the overall morphology of the central density profile.  
Annihilation alone introduces a characteristic density floor $\rho_{\rm core}$, while the companion injects additional energy and effectively increases the size of this low-density region.  
Balancing Eqs.~(\ref{eq:rhocore}) and (\ref{eq:rscour}) yields the approximate scaling
\begin{equation}
r_{\rm scour}\propto
\left[
(\langle\sigma v\rangle\,t_{\rm bin})^{\frac{\gamma_{\rm sp}-2}{\gamma_{\rm sp}}}
\,m_2^{3/2}\,m_1^{-1}\,r_2^{-1/2}\,
\rho_R^{-1}R_{\rm sp}^{-\gamma_{\rm sp}}
\right]^{1/(2-\gamma_{\rm sp})}.
\label{eq:rscaling}
\end{equation}
Thus, more massive and long-lived companions, shallower spikes, and smaller annihilation cross sections all enhance $r_{\rm scour}$ and the resulting suppression of the $J$ factor.

\textcolor{black}{In the canonical Gondolo--Silk case ($\gamma_{\rm sp}\simeq2.25$), the scaling exponent $(2-\gamma_{\rm sp})^{-1}\simeq-4$ makes $r_{\rm scour}$ only weakly sensitive to the companion's parameters.  However, because Eq.~(\ref{eq:Jratio_gammasp_gt_threehalves}) governs the $J$-factor for $\gamma_{\rm sp}>3/2$, the exponent $(3-2\gamma_{\rm sp})\approx-1.5$ means that even a moderate scouring ratio $r_{\rm scour}/r_{\rm core}\sim\mathrm{few}$ can suppress $J_{\rm mod}/J_{\rm sp}$ by an order of magnitude.  Whether $J_{\rm mod}/J_{\rm sp}\approx1$ or $J_{\rm mod}/J_{\rm sp}\ll1$ in the Gondolo--Silk regime therefore depends on whether the companion is sufficiently massive and long-lived to drive $r_{\rm scour}$ appreciably above $r_{\rm core}$.  For $\gamma_{\rm sp}\lesssim1.8$, the same dynamical energy drives $r_{\rm scour}$ to much larger multiples of $r_{\rm core}$, so the suppression condition is met even for modest companions---as confirmed by Figs.~\ref{fig:J_gamma} and~\ref{fig:heatmap}.}

\subsection{Physical interpretation}

This transition can be interpreted as the boundary between the ``annihilation-limited'' and ``dynamical-heating-limited'' regimes:
\begin{itemize}
\item In the annihilation-limited regime ($\gamma_{\rm sp}>2$), the DM distribution is so tightly bound that the annihilation plateau determines the inner cutoff.  The dark companion cannot easily overcome the steep potential barrier.
\item In the heating-limited regime ($\gamma_{\rm sp}<2$), the spike is already partially relaxed.  The binding energy per unit mass decreases toward small radii, allowing the same orbital energy to excavate a much larger region.
\end{itemize}
This dichotomy explains why only systems that have undergone significant stellar heating or past merger activity are expected to exhibit strong suppression of annihilation signals due to dark companions.  
The qualitative trend mirrors that found in simulations of stellar-density ``core scouring'' by binary black holes in galactic nuclei \cite{Merritt:2006mt,Gualandris:2010,Thomas:2014,Bortolas:2016}, underscoring the universality of the underlying dynamics.

\subsection{Implications for indirect detection}

The sensitivity of $J$ to $\gamma_{\rm sp}$ and to the effective core size implies that astrophysical modeling of the inner halo is as important as particle-physics uncertainties in $\langle\sigma v\rangle$.  
For instance, in a relaxed or cored halo with $\gamma_{\rm sp}\sim1.5$, the same annihilation cross section that would saturate current \emph{Fermi}-LAT limits under an NFW assumption may yield fluxes two orders of magnitude smaller.  
Neglecting the potential role of a companion in such environments could therefore lead to an overestimation of the expected annihilation signal and a corresponding underestimation of the required cross section.  

The next section quantifies these dependencies through explicit numerical integrations of the scoured density profile across the parameter ranges identified in Section~\ref{sec:constraints}.

\section{Observational Context and Constraints}
\label{sec:obs}

The inner density structure of the Milky Way’s dark-matter halo remains one of the most uncertain components of indirect detection modeling.  While cosmological $N$-body simulations predict a cuspy central profile, observations of stellar kinematics, gas dynamics, and gamma-ray emission allow for a broad range of effective inner slopes.  The true configuration of the Galactic dark-matter distribution within the central kiloparsec is likely shaped by the interplay of baryonic processes, dynamical heating, and possible dark-sector interactions.  Understanding these factors is essential for evaluating whether a dark companion could plausibly account for a suppressed annihilation signal from the Galactic Center (GC).

\subsection{Constraints from stellar and gas dynamics}

Stellar-dynamical modeling of the Galactic bulge and nuclear star cluster provides some of the tightest empirical bounds on the enclosed mass profile within $\sim1$\,kpc.  
Kinematic data from infrared integral-field spectroscopy and proper-motion surveys \cite{Feldmeier-Krause:2017,Schodel:2014,Ciambur:2017} are well described by mass distributions consistent with a near-isothermal or mildly cuspy DM halo, $\rho_\chi\propto r^{-\gamma}$ with $\gamma\approx0.6$–$1.2$.  
Within the central few parsecs, the gravitational potential is dominated by the stellar cluster and the supermassive black hole Sgr~A$^*$, leaving the DM contribution dynamically subdominant and largely unconstrained \cite{Schodel:2009,Gallego-Cano:2018}.  
Constraints from the stellar orbital distribution thus exclude neither a shallow core nor a steep spike below $\sim0.1$\,pc.

Gas dynamics and maser rotation curves in the inner bulge similarly indicate that baryonic components dominate the gravitational potential to $\sim500$\,pc, but they do not preclude a denser dark component embedded within the stellar cusp \cite{Sofue:2013,Launhardt:2002}.  
In this regime, uncertainties in the mass-to-light ratio of the nuclear stellar population translate directly into uncertainties in the inferred DM density at the $\sim$\,factor-of-3 level.

\subsection{Microlensing and dynamical heating constraints}

Microlensing surveys toward the Galactic bulge provide complementary limits on the fraction of non-luminous mass along the line of sight \cite{Wyrzykowski:2015,Bennett:2019}.  
These data disfavour strongly cusped halos within the inner few hundred parsecs, as such models would overpredict the optical depth of microlensing events.  
However, the constraints depend sensitively on the assumed stellar mass function and bar geometry; cored and moderately cusped profiles both remain viable \cite{Portail:2017}.  

Dynamical heating by baryonic substructures—stellar clusters, giant molecular clouds, and black holes—can also erode DM cusps over cosmic time \cite{El-Zant:2001,Governato:2012fa}.  
Simulations of the nuclear star cluster show that relaxation times are short enough to isotropize the phase-space distribution of dark matter within $\sim1$\,pc \cite{Merritt:2004xa,Vasiliev:2007zx}, potentially flattening any primordial spike.  
If a dark companion exists, its gravitational influence would amplify this heating and could account for a more pronounced central deficit in DM than would be expected from stellar interactions alone.

\subsection{Gamma-ray observations and the Galactic Center excess}

Gamma-ray data provide indirect but highly informative probes of the inner halo.  
Analyses of \emph{Fermi}-LAT observations have consistently revealed an extended emission component around the GC, peaking at a few GeV and approximately spherically symmetric \cite{Daylan:2014rsa,Calore:2015oya,DiMauro:2021raz}.  
If interpreted as DM annihilation, the signal favors an NFW-like profile with $\gamma\approx1.2$ and a total $J$ factor consistent with canonical expectations.  
However, recent morphological and spectral analyses suggest that the emission is likely dominated by unresolved point sources, such as millisecond pulsars \cite{Lee:2015fea,Bartels:2018}.  
If this interpretation is correct, the lack of an excess attributable to DM annihilation implies that the true inner $J$ factor may be substantially smaller than the canonical value—possibly by two or more orders of magnitude.

At higher energies, observations by H.E.S.S. and MAGIC have constrained TeV-scale emission consistent with the presence of cosmic-ray acceleration near Sgr~A$^*$ but inconsistent with bright DM annihilation spikes \cite{Abdallah:2016ygi,Ahnen:2017}.  
These results disfavour models predicting extreme Gondolo–Silk spikes unless additional suppression mechanisms—such as dynamical scouring—are present.

\subsection{Linking observations to theoretical expectations}

Taken together, existing constraints admit a wide range of possible inner slopes, from $\gamma\simeq0.6$ (cored or baryon-heated profiles) to $\gamma_{\rm sp}\simeq2.3$ (adiabatically contracted or spiked halos).  
This broad interval translates into variations of several orders of magnitude in the expected $J$ factor, underscoring the importance of incorporating dynamical effects.  

A dark companion orbiting Sgr~A$^*$ offers a natural astrophysical mechanism to reconcile the theoretical expectation of a steep spike with the apparent absence of a bright annihilation signal.  
By scouring the innermost region and effectively expanding the annihilation core, such a companion can suppress the gamma-ray flux without requiring modifications to the DM microphysics.  
As we demonstrate below, the magnitude of this suppression depends systematically on the spike slope, the companion’s mass and orbit, and the age of the binary system.

\section{Numerical Results and Parameter Exploration}
\label{sec:results}

We now quantify the suppression of the dark-matter annihilation $J$ factor due to a dark companion by numerically integrating the squared density along the line of sight and over a finite observation cone.  The calculations employ the formalism described in Sections~\ref{sec:scour} and~\ref{sec:spikecore}, using the full three-dimensional profiles rather than the simplified analytic approximations.  All integrations were performed with adaptive numerical routines that ensure convergence of the inner contributions, with a minimum softening radius $r_{\rm soft} = \max(5\,R_{\rm S}, 10^{-2}R_{\rm core})$ to avoid artificial divergences at $r\to0$.  { The dark-matter particle parameters are fixed at $m_\chi = 100~\mathrm{GeV}$ and
$\langle\sigma v\rangle = 10^{-26}\,\mathrm{cm}^3\,\mathrm{s}^{-1}$, with a system age
$t_{\rm age} = 1\text{--}5~\mathrm{Gyr}$ that sets the annihilation plateau.
We vary the companion’s mass $m_2$, orbital separation $r_2$, and binary lifetime
$t_{\rm bin}$ within the astrophysically allowed ranges.

Throughout this section we distinguish between the system age $t_{\rm age}$,
which determines the annihilation plateau and $R_{\rm core}$, and the binary
lifetime $t_{\rm bin}$, which controls the cumulative dynamical-friction energy
injected by the companion.


}

\subsection{Parameter space and benchmark configurations}

Our fiducial model adopts a supermassive black hole mass $m_1 = 4\times10^6\,M_\odot$ and a canonical halo normalization $\rho_0 = 3\times10^4\,M_\odot\,\mathrm{pc^{-3}}$ at $r_0 = 0.3\,\mathrm{pc}$, consistent with the inner bulge mass distribution.  {Specifically, we parametrize the underlying dark-matter halo profile in the vicinity of the
central black hole as a power law,
\begin{equation}
\rho_{\rm halo}(r)
=
\rho_0 \left( \frac{r}{r_0} \right)^{-\gamma},
\label{eq:halo_profile}
\end{equation}
where $\rho_0$ is the normalization of the pre-spike halo density at the
reference radius $r_0$, and $\gamma$ is the inner logarithmic slope of the halo
prior to spike formation. The parameters $\rho_0$ and $r_0$ are fixed by matching
to the large-scale halo profile and serve only as normalization constants in
the spike construction. Note that the parameters $\rho_0$ and $r_0$ affect the spike normalization but do not
introduce additional free physics beyond the choice of the underlying halo
profile.

}
The dark-matter particle parameters are fixed at $m_\chi = 100\,\mathrm{GeV}$ and $\langle\sigma v\rangle = 10^{-26}\,\mathrm{cm^3\,s^{-1}}$, with a system age $t_{\rm age} = 1$–$5\,\mathrm{Gyr}$.
We vary the companion’s mass $m_2$, orbital separation $r_2$, and lifetime $t_{\rm bin}$ within the astrophysically allowed ranges identified in Section~\ref{sec:constraints}:
\[
10^2\,M_\odot \le m_2 \le 10^5\,M_\odot, \quad
10\,\mathrm{AU} \le r_2 \le 10^4\,\mathrm{AU}, \quad
0.1\,\mathrm{Gyr} \le t_{\rm age} \le 10\,\mathrm{Gyr}.
\]
The spike slope $\gamma_{\rm sp}$ is varied between 1.5 and 2.4 to capture both relaxed and adiabatically contracted scenarios.

For each configuration, we compute:
\[
J_{\rm sp} = 
\int_{\mathrm{l.o.s.}}\rho_{\rm sp}^2(r[s])\,ds, \qquad
J_{\rm mod} = 
\int_{\mathrm{l.o.s.}}\rho_{\rm mod}^2(r[s])\,ds,
\]
and evaluate the suppression ratio $J_{\rm mod}/J_{\rm sp}$ and the dimensionless scouring parameter $s=r_{\rm scour}/R_{\rm core}$.

\subsection{Dependence on companion parameters}

\begin{figure}[t]
\centering
\fbox{\includegraphics[width=0.9\textwidth]{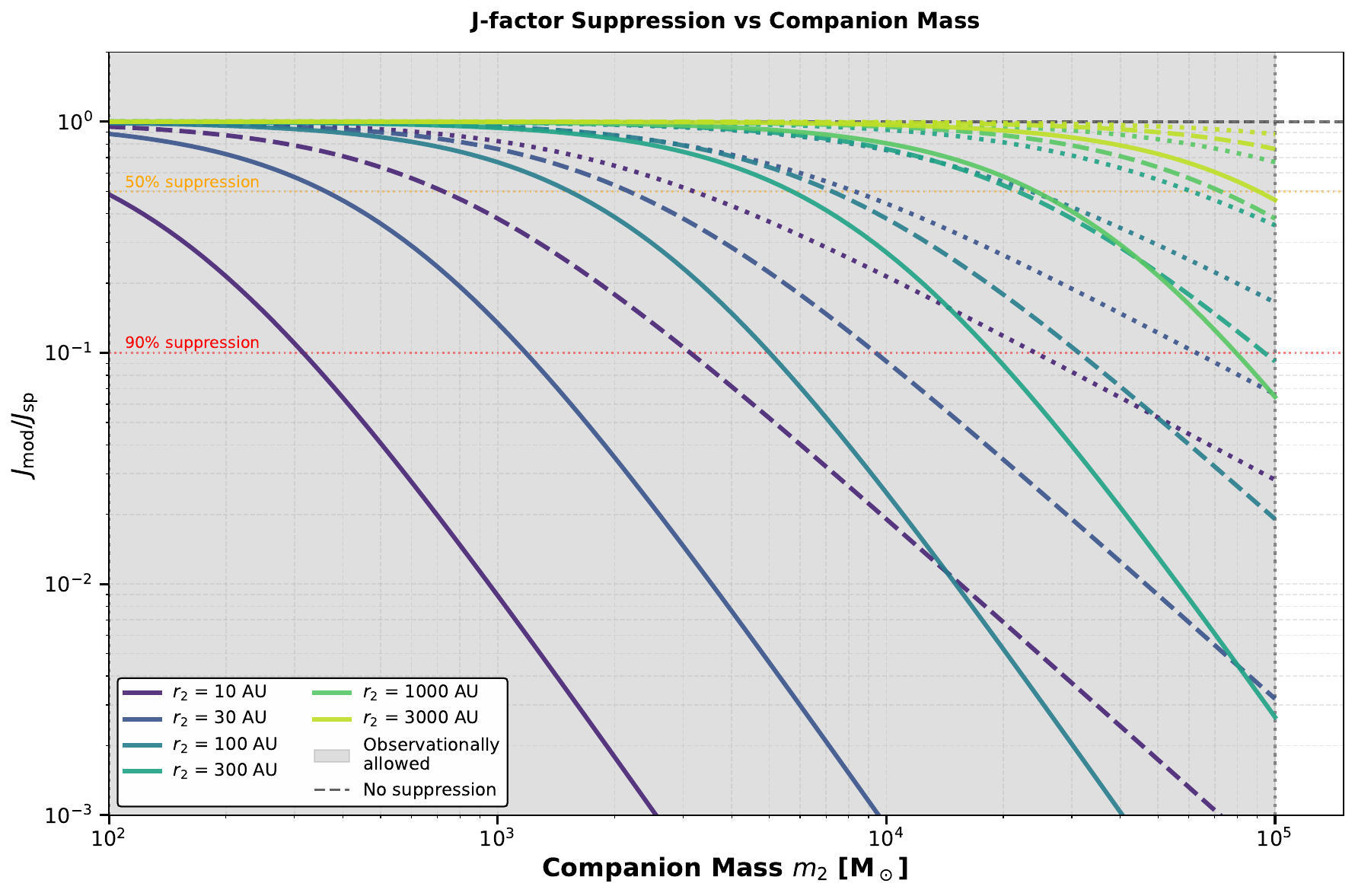}}
\caption{Suppression of the annihilation $J$ factor as a function of the companion mass $m_2$ for several orbital separations $r_2$.  
Solid, dashed, and dotted lines correspond to $\gamma_{\rm sp}=2.3,\,2.0,\,1.8$, respectively.  
The shaded band indicates the observationally allowed range of $m_2$ (see Fig.~\ref{fig:constraints}).}
\label{fig:J_m2_r2}
\end{figure}

Figure~\ref{fig:J_m2_r2} shows the variation of $J_{\rm mod}/J_{\rm sp}$ as a function of the companion mass $m_2$ for several fixed orbital separations $r_2$.  
For low-mass companions ($m_2\lesssim10^3\,M_\odot$), the effect is negligible across most separations.  
Above $m_2\sim10^4\,M_\odot$, the suppression becomes noticeable, with $J_{\rm mod}/J_{\rm sp}\sim0.1$ for tight orbits ($r_2\lesssim100$\,AU) {in systems with $t_{\rm age} \gtrsim 1~\mathrm{Gyr}$ and long-lived companions
($t_{\rm bin} \gtrsim 1~\mathrm{Gyr}$).}

This scaling follows directly from the $\Delta E_{\rm DF}\propto m_2^{3/2}$ dependence discussed in Section~\ref{sec:scour}.

\subsection{Dependence on spike slope and core size}

\begin{figure}[t]
\centering
\fbox{\includegraphics[width=0.9\textwidth]{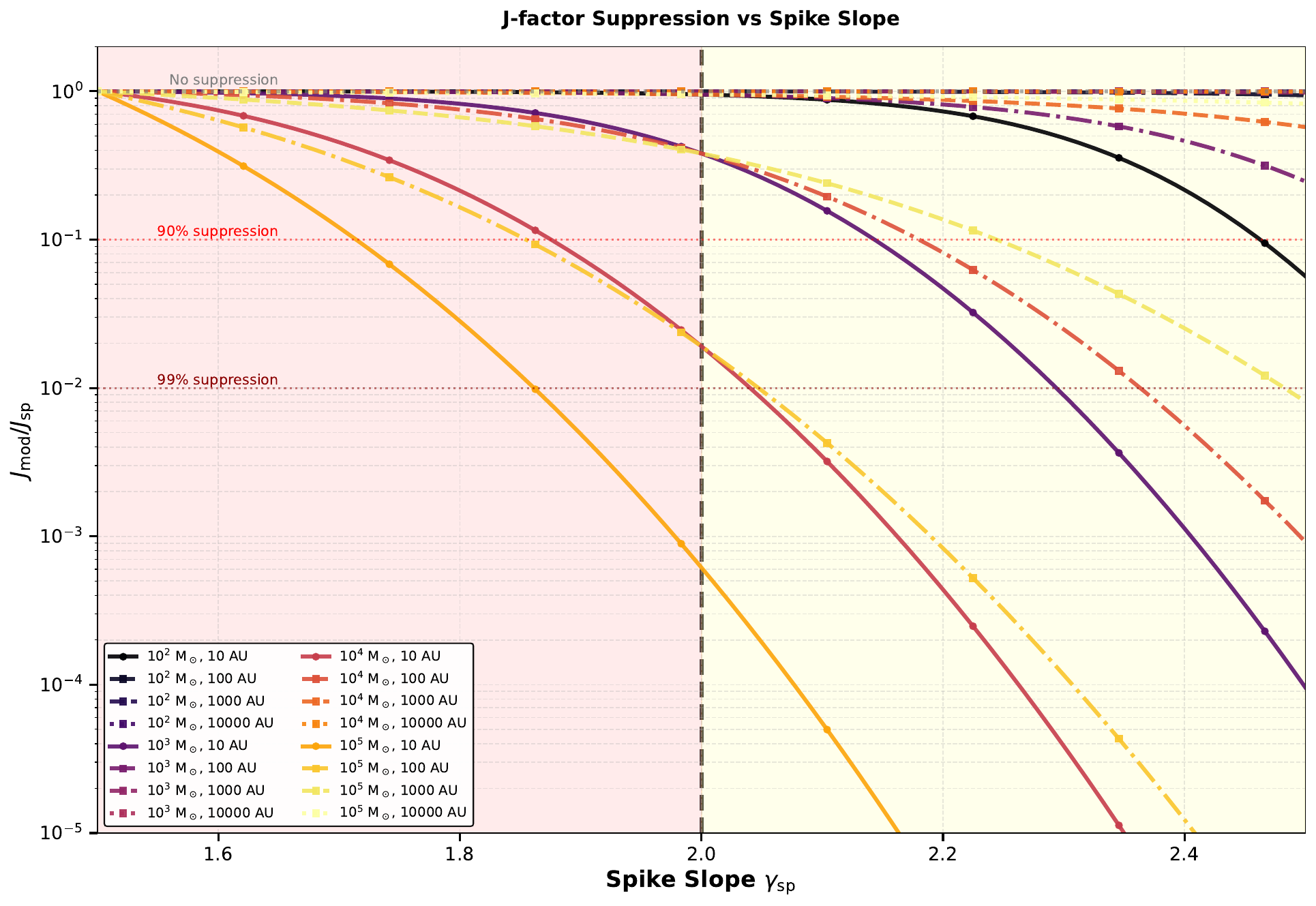}}
\caption{Dependence of the $J$-factor suppression $J_{\rm mod}/J_{\rm sp}$ on the spike slope $\gamma_{\rm sp}$, for representative values of $m_2$ and $r_2$.  
The vertical dashed line marks the transition between the Gondolo--Silk regime ($\gamma_{\rm sp}>2$) and the dynamically heated regime ($\gamma_{\rm sp}<2$).}
\label{fig:J_gamma}
\end{figure}

The dependence on $\gamma_{\rm sp}$ is illustrated in Figure~\ref{fig:J_gamma}.  
\textcolor{black}{For $\gamma_{\rm sp}>2$, the scouring radius grows only slowly with $\Delta E_{\rm DF}$, so significant excavation requires more massive or longer-lived companions; nevertheless, the steep $J$-factor scaling of Eq.~(\ref{eq:Jratio_gammasp_gt_threehalves}) means that once $r_{\rm scour}/r_{\rm core}$ exceeds a moderate value, the suppression can be substantial even in this regime.}
As $\gamma_{\rm sp}$ decreases below 2, however, the suppression becomes increasingly pronounced already for modest companions.  
For $\gamma_{\rm sp}=1.6$ and $m_2=10^4\,M_\odot$, the reduction can reach two orders of magnitude.  
This behavior confirms the analytical expectation that shallower spikes are dynamically fragile and more susceptible to scouring.

\subsection{Scaling with the scouring parameter $s$}

To test the analytical scaling relations, we evaluate $J_{\rm mod}/J_{\rm sp}$ as a function of $s = r_{\rm scour}/R_{\rm core}$.  
\textcolor{black}{Using the piecewise approximation---Eq.~(\ref{eq:Jmod_valid}) for $\gamma_{\rm sp}<3/2$ and Eq.~(\ref{eq:Jratio_gammasp_gt_threehalves}) for $\gamma_{\rm sp}\ge3/2$---we estimate the suppression of the annihilation signal as a function of the scouring radius.}
The log--log slopes of these relations follow the predicted power-law scalings derived in each regime.  
This correspondence validates the analytical framework of Section~\ref{sec:scour} and provides a convenient way to estimate the suppression for arbitrary spike parameters.

\begin{figure}[t]
\centering
\fbox{\includegraphics[width=0.9\textwidth]{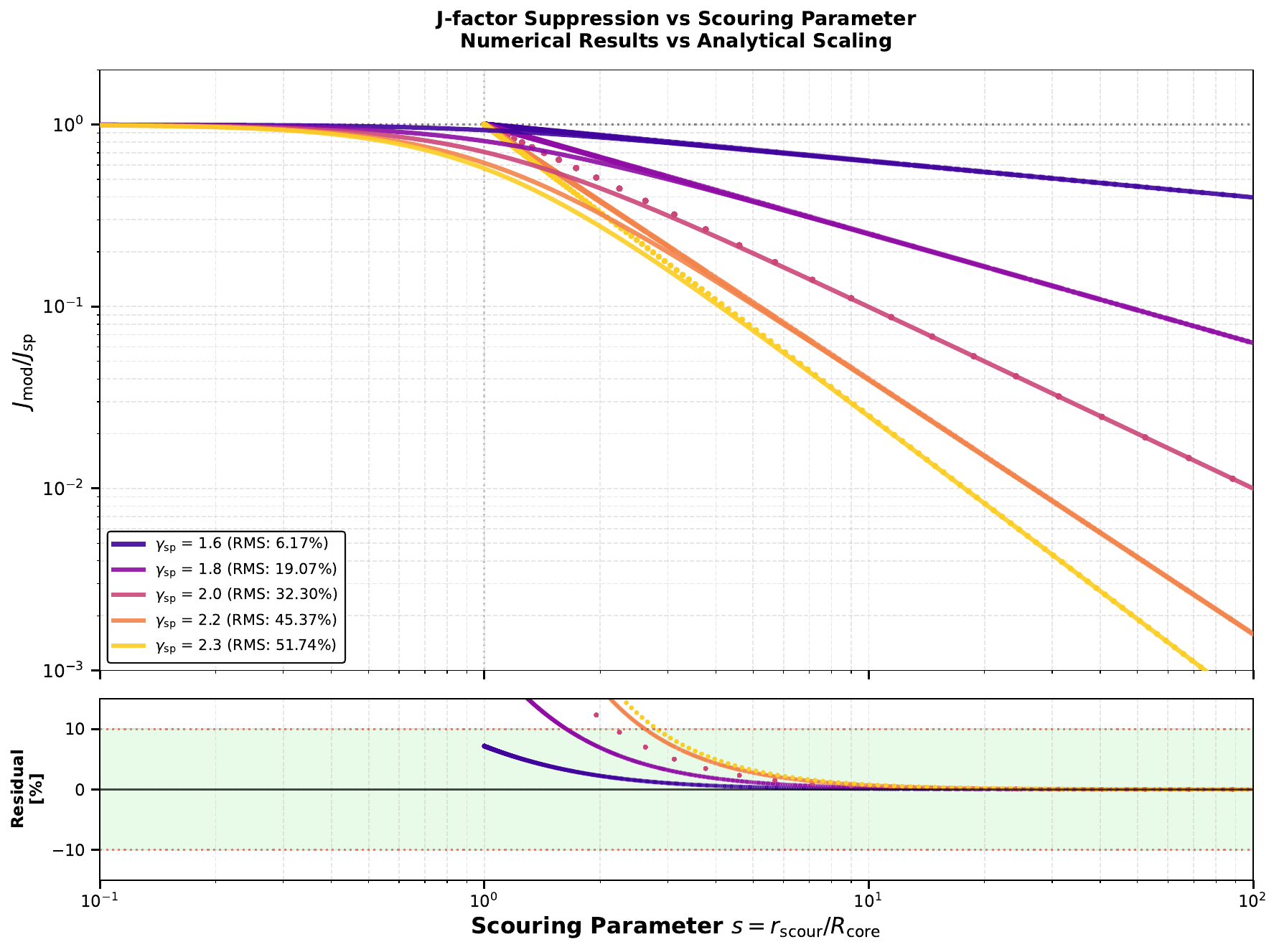}}
\caption{\textcolor{black}{Validation of the piecewise analytical approximation (Eq.~(\ref{eq:Jmod_valid}) for $\gamma_{\rm sp}<3/2$; Eq.~(\ref{eq:Jratio_gammasp_gt_threehalves}) for $\gamma_{\rm sp}\ge3/2$) against full numerical line-of-sight integrations (points) across multiple spike slopes. Agreement is excellent for all $\gamma_{\rm sp}$ values shown, confirming that the piecewise scaling captures the essential physics of $J$-factor suppression throughout the parameter space.}}
\label{fig:J_s}
\end{figure}

\subsection{Two-dimensional parameter scans}

\begin{figure*}[t]
\centering
\fbox{\includegraphics[width=0.9\textwidth]{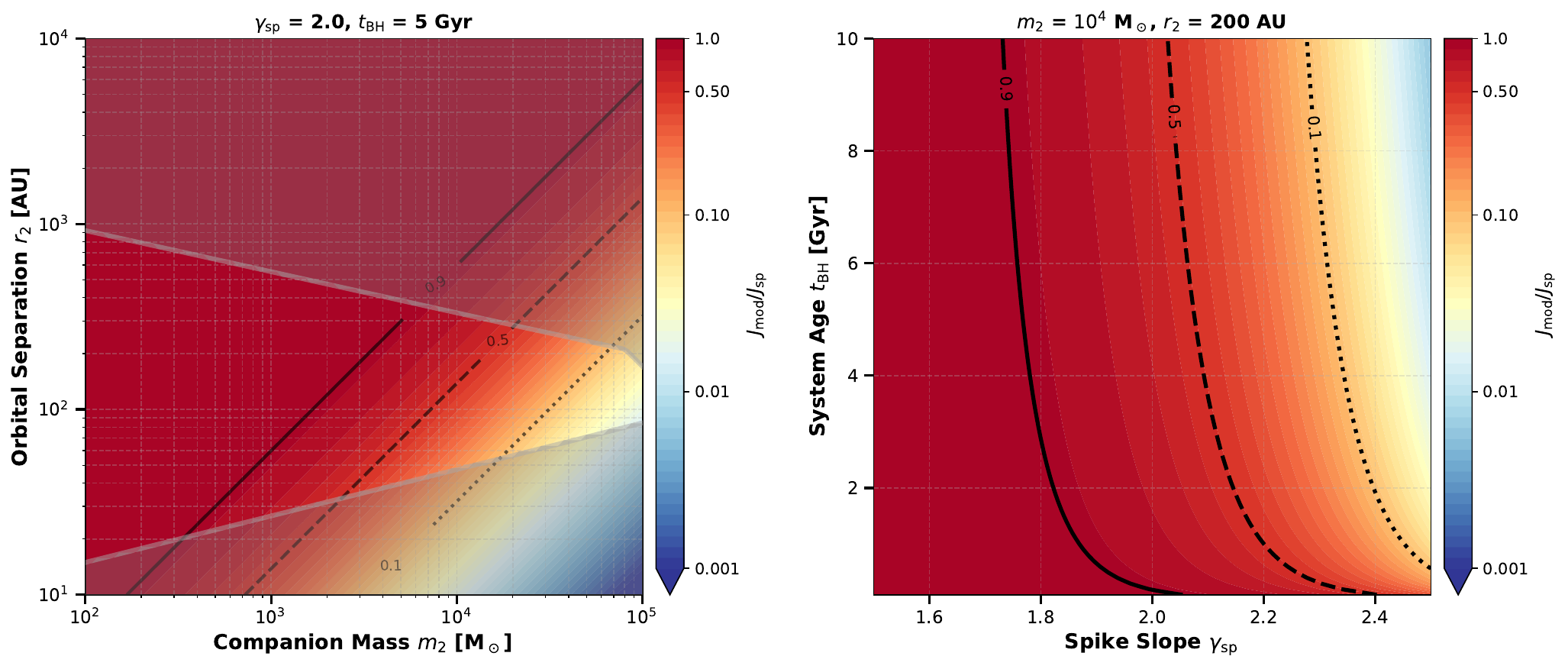}}
\caption{Parameter-space maps of the $J$-factor suppression ratio $J_{\rm mod}/J_{\rm sp}$.  
\emph{Left:} dependence on companion mass $m_2$ and orbital separation $r_2$ for $\gamma_{\rm sp}=2.0$ and $t_{\rm age}=t_{\rm bin}=5$\,Gyr.  
\emph{Right:} dependence on spike slope $\gamma_{\rm sp}$ and system age $t_{\rm age}$ for $m_2=10^4\,M_\odot$ and $r_2=200$\,AU.  
Contours mark suppression levels of $J_{\rm mod}/J_{\rm sp}=0.9,\,0.5,\,0.1$. The gray shaded region in the left panel represents the observationally excluded region from Figure
\ref{fig:constraints}.}
\label{fig:heatmap}
\end{figure*}

{ Figure~\ref{fig:heatmap} presents two-dimensional heatmaps of the suppression ratio.
In the right panel, we vary the system age $t_{\rm age}$, which controls the size
of the annihilation core $R_{\rm core}$, while holding the companion lifetime
$t_{\rm bin}$ fixed.
Longer system ages increase $R_{\rm core}$ and thus enhance the sensitivity
to scouring at fixed injected orbital energy. For simplicity, in this figure we set $t_{\rm bin}=t_{\rm age}$; relaxing this
assumption would rescale $r_{\rm scour}$ without modifying the qualitative trends.

These maps highlight the boundaries of the parameter space where scouring becomes observationally significant ($J_{\rm mod}/J_{\rm sp}\lesssim0.1$).  
The top-left panel shows that only relatively massive and compact companions ($m_2\gtrsim10^4\,M_\odot$, $r_2\lesssim10^2$\,AU) can measurably reduce the annihilation flux in a steep spike.  
The bottom-right panel demonstrates that longer system ages enhance suppression, consistent with the $\Delta E_{\rm DF}\propto t_{\rm bin}$ scaling.}

\subsection{Comparison with analytical expectations}

\begin{figure}[t]
\centering
\fbox{\includegraphics[width=0.9\textwidth]{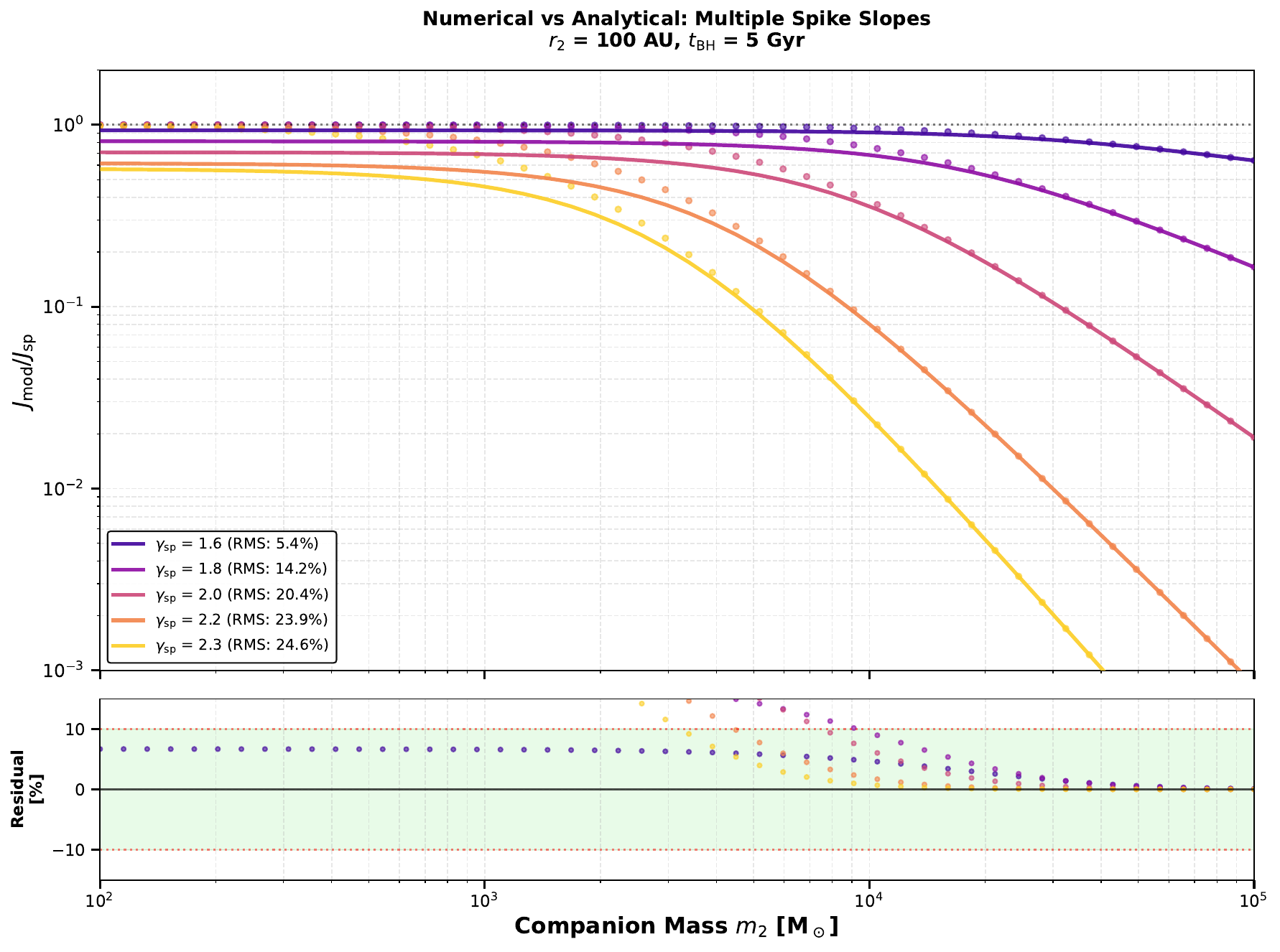}}
\caption{\textcolor{black}{Comparison between the full numerical results (points) and the piecewise analytical prediction---Eq.~(\ref{eq:Jmod_valid}) for $\gamma_{\rm sp}<3/2$ and Eq.~(\ref{eq:Jratio_gammasp_gt_threehalves}) for $\gamma_{\rm sp}\ge3/2$---}(solid lines). Residuals (lower panel) show deviations smaller than 10\% over the relevant parameter space.}
\label{fig:compare}
\end{figure}

To illustrate the consistency between the full numerical integrations and the simplified scouring-radius model, Figure~\ref{fig:compare} compares the numerically obtained $J_{\rm mod}/J_{\rm sp}$ values with \textcolor{black}{the piecewise analytical prescription---Eq.~(\ref{eq:Jmod_valid}) for $\gamma_{\rm sp}<3/2$ and Eq.~(\ref{eq:Jratio_gammasp_gt_threehalves}) for $\gamma_{\rm sp}\ge3/2$---using} the numerically computed $r_{\rm scour}$.  
The agreement is excellent (typically within $10\%$) across the entire physically relevant parameter range, confirming that the scouring-radius approximation captures the essential physics of the suppression.

\subsection{Summary of trends}

The principal trends emerging from the numerical exploration are:
\begin{enumerate}
\item \textcolor{black}{The $J$-factor suppression depends on the spike slope $\gamma_{\rm sp}$ through two competing effects: the scouring radius $r_{\rm scour}$ grows most easily for $\gamma_{\rm sp}<2$ (lower central binding energy), while for $\gamma_{\rm sp}>2$ a given $r_{\rm scour}/r_{\rm core}$ produces \emph{larger} fractional $J$-factor suppression via Eq.~(\ref{eq:Jratio_gammasp_gt_threehalves}).  In the shallow-spike regime ($\gamma_{\rm sp}\lesssim1.8$), even modest companions drive large scouring radii, yielding suppression of one to three orders of magnitude across much of the allowed parameter space.  In the steep-spike regime ($\gamma_{\rm sp}\gtrsim2$), the onset of significant suppression requires more massive or longer-lived companions, but once $r_{\rm scour}/r_{\rm core}\gg1$, the $J$-factor reduction can be equally severe.}
\item The companion’s mass is the dominant parameter controlling the onset of suppression, while the orbital separation determines its efficiency.
\item System age $t_{\rm age}$ acts linearly in the total deposited energy and thus in the logarithm of $J_{\rm mod}/J_{\rm sp}$.
\item The numerical data are well described by simple power-law fits in terms of the dimensionless parameter $s$, providing a practical prescription for indirect-detection forecasts.
\end{enumerate}

In the next section we discuss the astrophysical implications of these results for ongoing and future indirect searches, and how the presence of a dark companion modifies the interpretation of gamma-ray, neutrino, and antimatter constraints.


\section{Discussion and Implications}
\label{sec:discussion}

The results presented above demonstrate that the presence of a dark companion near
Sgr~A$^*$ fundamentally alters the astrophysical interpretation of indirect
dark-matter searches targeting the Galactic Center.  \textcolor{black}{Although steep Gondolo--Silk spikes are harder to scour---requiring
more massive or longer-lived companions to drive $r_{\rm scour}$ appreciably above $r_{\rm core}$---any
degree of prior heating or relaxation dramatically increases susceptibility to scouring,
and even steep spikes can exhibit strong suppression once scouring is initiated.}
In this section we synthesize these results and examine the broader implications
for gamma-ray, neutrino, and cosmic-ray searches, as well as for future
observations of the Galactic Center environment.

\subsection{Astrophysical degeneracies in the interpretation of indirect-detection data}

Indirect searches rely critically on the assumed inner-halo profile, and the
$J$-factor suppression induced by a dark companion introduces a major new
astrophysical uncertainty.  Current analyses of the Fermi-LAT Galactic Center
excess typically assume NFW-like cusps with $\gamma\approx1.2$ or, in some
cases, mildly contracted profiles, and find a reasonable fit to the data under a
DM interpretation~\cite{Daylan:2014rsa,Calore:2015oya,DiMauro:2021raz,Hooper:2013rwa}.  
However, population-studies and non-Poissonian template analyses suggest that
unresolved millisecond pulsars are likely to dominate the GeV emission
\cite{Lee:2015fea,Bartels:2018}, weakening the case for a bright annihilation
signal from the Galactic Center.  

Our results show that even a modest companion with
$m_2\sim10^4\,M_\odot$ on a $\sim100$--$300$ AU orbit can suppress the
annihilation flux by one to two orders of magnitude if the underlying spike is
shallower than $\gamma_{\rm sp}\simeq2$.  
The resulting degeneracy is severe: a null detection could be due to a small
annihilation cross section, a cored or dynamically heated halo
\cite{Pontzen:2011ty,Governato:2012fa,Chan:2015tna,Merritt:2004xa}, or an
otherwise steep cusp that has been partially excavated by a compact companion.
In other words, the absence of a bright signal near Sgr~A$^*$ does not uniquely
point to particle-physics explanations; it can equally well reflect the dynamical
history of the Galactic nucleus.

The situation is even more acute for TeV-scale gamma-ray searches such as
H.E.S.S.\ and CTA~\cite{Abdallah:2016ygi,CTAConsortium:2018tzg}, which are
particularly sensitive to the very inner region of the spike.  
A system that would naively predict a detectable signal under a
Gondolo--Silk assumption may fall entirely below existing limits once scouring
is accounted for.  
This implies that constraints on $\langle\sigma v\rangle$ in the TeV range may
currently be overly aggressive by up to an order of magnitude, depending on the
degree of heating or scouring the spike has experienced.

\subsection{Comparison with N-body simulations}

During the preparation of this work, a complementary study based on N-body simulations
appeared in Ref.~\cite{NBodySpike2025}, which investigates the modification of dark-matter
spikes in the presence of a massive companion through direct phase-space evolution.
While the physical setups are not identical, the two approaches address the same underlying
question: how dynamical heating by a bound companion suppresses the annihilation signal from
a central spike.

Ref.~\cite{NBodySpike2025} follows the time-dependent redistribution of dark-matter particles
induced by the binary potential, whereas our analysis adopts a semi-analytic energy-balance
criterion to estimate an effective scouring radius and computes the resulting $J$-factor
suppression through line-of-sight integrations of the modified density profile.
As a result, quantitative differences are expected, particularly in how sharply the inner
profile is flattened and how the transition region around $r_{\rm scour}$ is modeled.

Once the slope-dependent scaling of the annihilation integral is treated consistently,
however, the qualitative behavior is in good agreement.
For $\gamma_{\rm sp}<3/2$, both approaches find that the annihilation signal is dominated by
the outer spike and that removing only the inner region produces a modest suppression when
$r_{\rm scour}\ll R_{\rm sp}$.
For $\gamma_{\rm sp}>3/2$, both approaches show that the suppression is controlled primarily
by the smallest surviving radius, leading to a steep dependence on the ratio
$r_{\rm scour}/r_{\rm core}$.

In the marginal case $\gamma_{\rm sp}=3/2$, Ref.~\cite{NBodySpike2025} reports a non-negligible
suppression even for moderate scouring.
This behavior is fully consistent with the logarithmic dependence derived in
Eq.~(\ref{eq:Jratio_gammasp_threehalves}), which predicts a finite reduction whenever
$r_{\rm scour}>r_{\rm core}$, despite the vanishing of the corresponding power-law exponent.
Any apparent disagreement therefore arises from treating the
$\gamma_{\rm sp}=3/2$ case as a strict boundary between regimes rather than as a distinct
logarithmic limit.

Residual quantitative differences between the two studies likely reflect the different
levels of modeling detail: the N-body simulations naturally capture gradual profile
flattening and time-dependent relaxation effects, while our approach provides a transparent,
computationally inexpensive framework that isolates the dominant parametric dependencies.
Taken together, the two methods are complementary, and their overall consistency strengthens
the conclusion that even modest dynamical heating can significantly alter the indirect-detection
prospects for dark-matter spikes near the Galactic Center.

\subsection{Complementarity with other spike-eroding mechanisms}

Dark companions are not the only agents capable of reshaping the inner
dark-matter profile.  
Stellar encounters, stellar-mass black-hole populations, molecular-cloud
passages, and past major or minor mergers all contribute to the erosion of
steep spikes over Gyr timescales
\cite{Merritt:2002vj,Gnedin:2003rj,Vasiliev:2007zx,Merritt:2006mt,Gualandris:2010,Thomas:2014,Bortolas:2016}.  
The key difference is that the scouring process analyzed here operates
\emph{coherently}, depositing energy through the continuous orbital motion of a
single massive body.  
This produces a more localized and often stronger effect near the annihilation
core than stochastic relaxation processes, especially in the radial range that
dominates the $J$ factor.

In fact, the scouring mechanism is highly complementary to relaxation-driven
heating:
\begin{itemize}
\item Stellar heating and merger activity lower $\gamma_{\rm sp}$, making the spike dynamically fragile and pushing it into the $\gamma_{\rm sp}\lesssim2$ regime \cite{Merritt:2004xa,Vasiliev:2007zx}.
\item A dark companion then amplifies this fragility, rapidly increasing
      $r_{\rm scour}/R_{\rm core}$ and driving a substantial reduction of the inner density.
\end{itemize}
This synergy means that even a mildly heated spike (e.g.\ $\gamma_{\rm sp}\sim1.8$)
could be nearly erased in the presence of a $10^4$--$10^5\,M_\odot$ companion,
whereas the same companion would have negligible effect on an adiabatic spike.

\subsection{Implications for interpreting Galactic Center observations}

Several observational puzzles may find a natural explanation within this
framework:
\begin{enumerate}
\item \textbf{The lack of a bright annihilation spike in TeV gamma rays.}  
      H.E.S.S.\ observations show no evidence of a sharply peaked annihilation
      component near Sgr~A$^*$, even though a canonical spike could easily
      overshoot existing limits~\cite{Abdallah:2016ygi}.  
      A dark companion provides a concrete dynamical mechanism for suppressing
      this emission without invoking exotic particle physics.

\item \textbf{The spectral and morphological properties of the Fermi-LAT Galactic Center excess.}  
      If unresolved millisecond pulsars dominate the GeV emission
      \cite{Lee:2015fea,Bartels:2018}, the absence of a DM-induced component may
      be due not only to particle physics but also to the astrophysical
      suppression described here.  
      The combined effect of heating and scouring can make an underlying DM
      contribution too faint to be separately identified, even if the Milky Way
      hosted a spike at earlier times.

\item \textbf{The inferred bounds on $\langle\sigma v\rangle$.}  
      Constraints assuming an unperturbed inner cusp or spike
      \cite{Hooper:2013rwa,Abazajian:2020tww,Karwin:2021jzp,CTAConsortium:2018tzg}
      may be biased low by factors of $10$--$100$ if the Milky Way hosts a
      dynamically relevant companion.  
      Future analyses should incorporate scouring-based priors when interpreting
      line-of-sight fluxes from the Galactic Center.
\end{enumerate}

\subsection{Prospects for multimessenger signatures of a hidden companion}

The same companion responsible for suppressing the annihilation signal may be
accessible to direct observation.  Several multimessenger channels are
particularly promising:
\begin{itemize}
\item \textbf{Astrometric signatures:}  
      Long-term monitoring of S-star orbits or the reflex motion of Sgr~A$^*$
      already constrains large parts of the $(m_2,a_2)$ plane
      \cite{Ghez:1998,Gravity:2020,Naoz:2019,Reid:2004,EHT:2022}.  Improved
      astrometric precision and extended time baselines could reveal
      $\sim10^3$--$10^5\,M_\odot$ objects at distances of $10$--$10^4$ AU.

\item \textbf{Low-frequency gravitational waves:}  
      A $10^3$--$10^5\,M_\odot$ compact object orbiting Sgr~A$^*$ lies in the
      sensitivity band of space-based interferometers such as LISA
      \cite{LISA:2017}.  
      If detected, such a source would directly measure the energy reservoir
      available to reshape the spike.

\item \textbf{Perturbations of the accretion flow:}  
      A companion embedded in the accretion environment of Sgr~A$^*$ could
      imprint characteristic variability or lensing signatures in mm/sub-mm
      emission accessible to VLBI facilities such as the EHT
      \cite{EHT:2022,Gillessen:2017}.
\end{itemize}
Any such detection would dramatically refine the astrophysical modeling
presented here and allow for a more precise evaluation of the $J$-factor
suppression.

\subsection{Implications for future Galactic Center surveys}

Upcoming gamma-ray and neutrino observatories will significantly improve
sensitivity to annihilating dark matter.  CTA will extend the dynamic range and
angular resolution of current Cherenkov arrays \cite{CTAConsortium:2018tzg},
while wide-field facilities such as SWGO will provide continuous monitoring of
the southern sky, including the Galactic Center
\cite{SWGO:2019WhitePaper}.  
In the neutrino sector, IceCube-Gen2 and KM3NeT will greatly enhance the reach
of existing telescopes~\cite{Abbasi:2021xzx,Adrian-Martinez:2016fei,IceCubeGen2:2023,KM3NeT:2016LoI}.  
Our results suggest that interpreting the data from these instruments will
require careful marginalization over the possible presence of a dark companion
and over the structured prior on $\gamma_{\rm sp}$ informed by baryonic physics
and nuclear star-cluster dynamics
\cite{Merritt:2004xa,Vasiliev:2007zx,Schodel:2014,Feldmeier-Krause:2017}.

Conversely, improved dynamical modeling of the Galactic Center---for example
through high-precision astrometry and spectroscopy of the nuclear star cluster
\cite{Schodel:2009,Gallego-Cano:2018,Feldmeier-Krause:2017} and continued EHT
observations of Sgr~A$^*$~\cite{EHT:2022}---will progressively shrink the
allowed $(m_2,r_2)$ parameter space and thus reduce the astrophysical
uncertainty in the $J$ factor.  
A joint analysis combining gamma-ray, neutrino, and gravitational-wave data
with stellar-dynamical constraints could, in principle, break the degeneracy
between particle physics and astrophysical suppression mechanisms.

\subsection{A revised view of the ``Galactic Center laboratory''}

The standard paradigm treats the inner parsec of the Milky Way as a fixed
environment for dark-matter annihilation searches, with uncertainties dominated
by baryonic modeling.  
The results of this work instead suggest a dynamical framework in which the
central density profile is shaped by the interplay of:
\begin{enumerate}
\item the historical assembly of the nuclear star cluster,
\item the adiabatic or non-adiabatic growth of Sgr~A$^*$,
\item stochastic stellar heating,
\item possible merger activity, and
\item the presence of a compact companion.
\end{enumerate}
In such a framework, the Milky Way’s inner halo is not a static target but a
time-evolving system whose structure retains memory of its dynamical history.  
The suppression of the annihilation signal by a compact companion is therefore
not an exotic possibility but a natural consequence of hierarchical galaxy
formation~\cite{Kormendy:1995yz,Lotz:2011av,Unavane:1996}.

\subsection{Summary of key implications}

Our main findings can be distilled into the following points:
\begin{itemize}
\item \textcolor{black}{A dark companion can suppress the annihilation signal by factors of
      $10$--$100$ or more across a broad range of spike slopes.  The effect is easiest to achieve for shallow spikes ($\gamma_{\rm sp}\lesssim2$), where even modest companions drive $r_{\rm scour}\gg R_{\rm core}$; for steep spikes ($\gamma_{\rm sp}\gtrsim2$), comparable suppression requires more massive or longer-lived companions, but is still achievable within the astrophysically allowed parameter space.}
\item Steep, adiabatically formed spikes are resilient; shallow, relaxed spikes
      are fragile and can be almost entirely erased.
\item The suppression depends primarily on the ratio $s=r_{\rm scour}/R_{\rm core}$,
      which allows a simple parametrization of the effect and matches the
      numerical results at the $\lesssim 10\%$ level.
\item Neglecting the possibility of a dark companion leads to overly strong
      constraints on $\langle\sigma v\rangle$, especially at the TeV scale where
      the spike contribution is most important.
\item Dynamical signatures, gravitational waves, and improved stellar
      astrometry offer promising ways to detect or constrain the companion
      responsible for scouring.
\end{itemize}

The next section summarizes the key results of this work and outlines future
directions for combining indirect detection, stellar dynamics, and
gravitational-wave observations to fully characterize the inner Galactic
dark-matter distribution.


\section{Conclusions}
\label{sec:conclusions}

The inner parsec of the Milky Way has long been regarded as one of the most
promising environments for detecting annihilating dark matter, owing to the high
density expected around the supermassive black hole Sgr~A$^*$.  The canonical
Gondolo--Silk spike, formed through adiabatic growth of the black hole, would
dramatically enhance the annihilation luminosity and render the Galactic Center
a uniquely bright target for indirect searches.  Yet the absence of a clear
annihilation signature in present gamma-ray, neutrino, and antimatter data
suggests that this idealized picture is incomplete.  In this work we have
demonstrated that the dynamical influence of a long-lived compact object---a
\emph{dark companion}---provides a robust and physically well-motivated
mechanism for suppressing the annihilation signal, even without invoking any
non-standard particle physics.

Our analysis integrates three key components: (i) a self-consistent description
of the unperturbed dark-matter profile, including spike formation,
annihilation-driven core development, and stellar heating; (ii) an analytic
framework that captures the dynamical impact of a secondary compact object
through a well-defined \emph{scouring radius}; and (iii) full numerical
evaluations of the line-of-sight integral for a wide range of dark companion
masses, orbital separations, system ages, and spike slopes.  These elements
combine to yield a clear physical picture: the extent to which a companion
suppresses the annihilation signal is controlled primarily by the ratio
$s=r_{\rm scour}/R_{\rm core}$, the companion’s mass and lifetime, and the slope
of the underlying spike.

Three main conclusions emerge from our study:

\begin{enumerate}
\item \textbf{Dynamically heated or relaxed spikes are highly susceptible to
      suppression.}  
      \textcolor{black}{When $\gamma_{\rm sp}\lesssim2$, the binding energy is lower at small radii and the companion can easily excavate a region significantly larger than the annihilation core; even a modest companion with $m_2\sim10^4\,M_\odot$ at separations of $\mathcal{O}(100\,\mathrm{AU})$ can reduce the annihilation flux by one to two orders of magnitude.  For $\gamma_{\rm sp}\gtrsim2$, the onset of large scouring requires more extreme binaries, but the steep $J$-factor scaling of Eq.~(\ref{eq:Jratio_gammasp_gt_threehalves}) means that comparable suppression levels are achievable once $r_{\rm scour}/r_{\rm core}$ exceeds a few.}  The effect is amplified when the nuclear star cluster has already softened the spike.

\item \textbf{Adiabatic Gondolo--Silk spikes are harder to scour but not
      immune to suppression.}  
      \textcolor{black}{For $\gamma_{\rm sp}>2$, the binding energy diverges toward the center and the scouring radius grows only weakly with the injected orbital energy, requiring more massive and long-lived companions to drive significant excavation.  However, once $r_{\rm scour}$ exceeds $r_{\rm core}$ by a moderate factor, the steep $J$-factor scaling $J_{\rm mod}/J_{\rm sp}\propto(r_{\rm scour}/r_{\rm core})^{3-2\gamma_{\rm sp}}$ can produce order-of-magnitude suppression even in this regime.}  Such steep spikes are in any case already in tension with gamma-ray observations from
      H.E.S.S.\ and with dynamical modeling of the nuclear star cluster, which
      point toward shallower present-day slopes.

\item \textbf{Astrophysical suppression creates major degeneracies in the
      interpretation of indirect searches.}  
      A null detection from the Galactic Center does not uniquely imply small
      annihilation cross sections; it may instead reflect the dynamical history
      of the nucleus.  Constraints on $\langle\sigma v\rangle$ that assume a
      smooth or mildly contracted halo may be overly stringent by factors of
      $10$--$100$ if even a single intermediate-mass companion is present.
      Future indirect searches must therefore marginalize over a structured
      astrophysical prior that includes dynamical scouring.
\end{enumerate}

These findings frame the Galactic Center not as a static annihilation
laboratory but as a dynamical environment whose structure reflects the
cumulative interplay of adiabatic contraction, stellar heating, black-hole
growth, and potentially the inspiral of a secondary compact object.  The
presence of such a companion is a natural outcome of hierarchical galaxy
formation, and future surveys will be increasingly sensitive to its
gravitational, astrometric, and gravitational-wave signatures.  Facilities such
as GRAVITY+, ngEHT, LISA, CTA, SWGO, and IceCube-Gen2 will probe the relevant
parameter space with unprecedented precision, offering multiple, complementary
pathways to detect or rule out dark companions in the $10^3$--$10^5\,M_\odot$
range.

Ultimately, our results emphasize that astrophysical uncertainties---and in
particular those associated with the dynamical state of the innermost halo---are
inseparable from particle-physics interpretations of indirect-detection data.
The scoured-spike framework developed here provides both a conceptual bridge and
a practical tool for connecting these perspectives.  A more complete
understanding of the Galactic Center will require a coordinated, multimessenger
approach combining gamma rays, neutrinos, gravitational waves, and precision
stellar dynamics.  Such an approach promises not only to clarify the role of
dark companions but also to sharpen the search for annihilating dark matter in
the most extreme and revealing environment in the Milky Way.

\section*{Acknowledgements}
We acknowledge significant early work on this project by Samuel D. English, Benjamin V. Lehmann, and Violet McAllister. SP is partly supported by the U.S. Department of Energy grant number de-sc0010107.

\bibliographystyle{apsrev4-2}
\bibliography{updatedrefs}

@article{Goodenough:2009gk,
  author = "Goodenough, Lisa and Hooper, Dan",
  title = "{Possible Evidence For Dark Matter Annihilation In The Inner Milky Way From The Fermi Gamma Ray Space Telescope}",
  eprint = "0910.2998",
  archivePrefix = "arXiv",
  primaryClass = "hep-ph",  
journal = "Unpublished",
  year = "2009"
}

@article{Daylan:2014rsa,
  author = "Daylan, Tansu and others",
  title = "{The characterization of the gamma-ray signal from the central Milky Way: A case for annihilating dark matter}",
  journal = "Phys. Dark Univ.",
  volume = "12",
  pages = "1--23",
  year = "2016",
  eprint = "1402.6703",
  archivePrefix = "arXiv",
  primaryClass = "astro-ph.HE"
}

@article{Calore:2015oya,
  author = "Calore, Francesca and Cholis, Ilias and Weniger, Christoph",
  title = "{Background model systematics for the Fermi GeV excess}",
  journal = "JCAP",
  volume = "1503",
  pages = "038",
  year = "2015",
  eprint = "1409.0042",
  archivePrefix = "arXiv",
  primaryClass = "astro-ph.CO"
}

@article{Ajello:2016sxc,
  author = "Ajello, M. and others",
  collaboration = "Fermi-LAT",
  title = "{Characterizing the population of pulsars in the Galactic bulge with the Fermi Large Area Telescope}",
  journal = "Astrophys. J.",
  volume = "819",
  number = "1",
  pages = "44",
  year = "2016",
  eprint = "1511.02938",
  archivePrefix = "arXiv",
  primaryClass = "astro-ph.HE"
}

@article{DiMauro:2021raz,
  author = "Di Mauro, Mattia",
  title = "{Characteristics of the Galactic Center excess measured with 11 years of Fermi-LAT data}",
  journal = "Phys. Rev. D",
  volume = "103",
  number = "6",
  pages = "063029",
  year = "2021",
  eprint = "2101.04694",
  archivePrefix = "arXiv",
  primaryClass = "astro-ph.HE"
}

@article{Bartels:2018qgr,
  author = "Bartels, Richard and Krishnamurthy, Shravan and Weniger, Christoph",
  title = "{Strong support for the millisecond pulsar origin of the Galactic center GeV excess}",
  journal = "Phys. Rev. Lett.",
  volume = "116",
  number = "5",
  pages = "051102",
  year = "2016",
  eprint = "1506.05104",
  archivePrefix = "arXiv",
  primaryClass = "astro-ph.HE"
}

@article{Leane:2019xiy,
  author = "Leane, Rebecca K. and Slatyer, Tracy R.",
  title = "{Spurious Signals of Dark Matter in the Galactic Center}",
  journal = "Phys. Rev. Lett.",
  volume = "123",
  number = "24",
  pages = "241101",
  year = "2019",
  eprint = "1904.08430",
  archivePrefix = "arXiv",
  primaryClass = "astro-ph.HE"
}

@article{Navarro:1996gj,
  author = "Navarro, Julio F. and Frenk, Carlos S. and White, Simon D. M.",
  title = "{The Structure of cold dark matter halos}",
  journal = "Astrophys. J.",
  volume = "462",
  pages = "563--575",
  year = "1996",
  eprint = "astro-ph/9508025"
}

@article{Navarro:2008kc,
  author = "Navarro, Julio F. and others",
  title = "{The Diversity and Similarity of Cold Dark Matter Halos}",
  journal = "Mon. Not. Roy. Astron. Soc.",
  volume = "402",
  pages = "21--34",
  year = "2010",
  eprint = "0810.1522",
  archivePrefix = "arXiv",
  primaryClass = "astro-ph"
}

@article{Springel:2008cc,
  author = "Springel, Volker and others",
  title = "{The Aquarius Project: the subhalos of galactic halos}",
  journal = "Mon. Not. Roy. Astron. Soc.",
  volume = "391",
  pages = "1685--1711",
  year = "2008",
  eprint = "0809.0898",
  archivePrefix = "arXiv",
  primaryClass = "astro-ph"
}

@article{Diemand:2008in,
    author = "Diemand, J. and Kuhlen, M. and Madau, P. and Zemp, M. and Moore, B. and Potter, D. and Stadel, J.",
    title = "{Clumps and streams in the local dark matter distribution}",
    eprint = "0805.1244",
    archivePrefix = "arXiv",
    primaryClass = "astro-ph",
    doi = "10.1038/nature07153",
    journal = "Nature",
    volume = "454",
    pages = "735--738",
    year = "2008"
}

@article{Blumenthal:1985qy,
       author = {{Blumenthal}, G.~R. and {Faber}, S.~M. and {Flores}, R. and {Primack}, J.~R.},
        title = "{Contraction of Dark Matter Galactic Halos Due to Baryonic Infall}",
      journal = {\apj},
         year = 1986,
        month = feb,
       volume = {301},
        pages = {27},
          doi = {10.1086/163867},
       adsurl = {https://ui.adsabs.harvard.edu/abs/1986ApJ...301...27B}
}

@article{Gnedin:2004cx,
  author = "Gnedin, Oleg Y. and Kravtsov, Andrey V. and Klypin, Anatoly A. and Nagai, Daisuke",
  title = "{Response of dark matter halos to condensation of baryons: Cosmological simulations and improved adiabatic contraction model}",
  journal = "Astrophys. J.",
  volume = "616",
  pages = "16--26",
  year = "2004",
  eprint = "astro-ph/0406247"
}

@article{Pontzen:2011ty,
  author = "Pontzen, Andrew and Governato, Fabio",
  title = "{How supernova feedback turns dark matter cusps into cores}",
  journal = "Mon. Not. Roy. Astron. Soc.",
  volume = "421",
  pages = "3464",
  year = "2012",
  eprint = "1106.0499",
  archivePrefix = "arXiv",
  primaryClass = "astro-ph.CO"
}

@article{Governato:2012fa,
  author = "Governato, Fabio and others",
  title = "{Cuspy no more: how outflows affect the central dark matter and baryon distribution in Lambda CDM galaxies}",
  journal = "Mon. Not. Roy. Astron. Soc.",
  volume = "422",
  pages = "1231--1240",
  year = "2012",
  eprint = "1202.0554",
  archivePrefix = "arXiv",
  primaryClass = "astro-ph.CO"
}

@article{Hooper:2013rwa,
  author = "Hooper, Dan and Slatyer, Tracy R.",
  title = "{Two Emission Mechanisms in the Fermi Bubbles: A Possible Signal of Annihilating Dark Matter}",
  journal = "Phys. Dark Univ.",
  volume = "2",
  pages = "118--138",
  year = "2013",
  eprint = "1302.6589",
  archivePrefix = "arXiv",
  primaryClass = "astro-ph.HE"
}

@article{Abazajian:2020tww,
  author = "Abazajian, Kevork N. and others",
  title = "{Strong Constraints on Thermal Relic Dark Matter from Fermi-LAT Observations of the Galactic Center}",
  journal = "Phys. Rev. D",
  volume = "102",
  number = "4",
  pages = "043012",
  year = "2020",
  eprint = "2003.10416",
  archivePrefix = "arXiv",
  primaryClass = "astro-ph.HE"
}

@article{Karwin:2021jzp,
    author = "Karwin, Christopher and Murgia, Simona and Tait, Tim M. P. and Porter, Troy A. and Tanedo, Philip",
    title = "{Dark Matter Interpretation of the Fermi-LAT Observation Toward the Galactic Center}",
    eprint = "1612.05687",
    archivePrefix = "arXiv",
    primaryClass = "hep-ph",
    reportNumber = "UCI-HEP-TR-2016-22",
    doi = "10.1103/PhysRevD.95.103005",
    journal = "Phys. Rev. D",
    volume = "95",
    number = "10",
    pages = "103005",
    year = "2017"
}

@article{Gondolo:1999ef,
  author = "Gondolo, Paolo and Silk, Joseph",
  title = "{Dark matter annihilation at the galactic center}",
  journal = "Phys. Rev. Lett.",
  volume = "83",
  pages = "1719--1722",
  year = "1999",
  eprint = "astro-ph/9906391"
}

@article{Merritt:2002vj,
  author = "Merritt, David and Milosavljevic, Milos and Verde, Licia and Jimenez, Raul",
  title = "{Dark matter spikes and annihilation radiation from the galactic center}",
  journal = "Phys. Rev. Lett.",
  volume = "88",
  pages = "191301",
  year = "2002",
  eprint = "astro-ph/0201376"
}

@article{Gnedin:2003rj,
  author = "Gnedin, Oleg Y. and Primack, Joel R.",
  title = "{Dark matter profile in the Galactic Center}",
  journal = "Phys. Rev. Lett.",
  volume = "93",
  pages = "061302",
  year = "2004",
  eprint = "astro-ph/0308385"
}

@article{Vasiliev:2007zx,
  author = "Vasiliev, Eugene",
  title = "{Dynamical evolution of dark matter spikes}",
  journal = "Phys. Rev. D",
  volume = "76",
  pages = "103532",
  year = "2007",
  eprint = "0707.3334",
  archivePrefix = "arXiv",
  primaryClass = "astro-ph"
}

@article{Fields:2014pia,
  author = "Fields, Brian D. and Shapiro, Stuart L. and Shelton, Jessie",
  title = "{Galactic Center Gamma-Ray Excess from Dark Matter Annihilation: Is There a Black Hole Spike?}",
  journal = "Phys. Rev. Lett.",
  volume = "113",
  number = "15",
  pages = "151302",
  year = "2014",
  eprint = "1406.4856",
  archivePrefix = "arXiv",
  primaryClass = "astro-ph.HE"
}

@article{Cirelli:2010xx,
    author = "Cirelli, Marco and Corcella, Gennaro and Hektor, Andi and Hutsi, Gert and Kadastik, Mario and Panci, Paolo and Raidal, Martti and Sala, Filippo and Strumia, Alessandro",
    title = "{PPPC 4 DM ID: A Poor Particle Physicist Cookbook for Dark Matter Indirect Detection}",
    eprint = "1012.4515",
    archivePrefix = "arXiv",
    primaryClass = "hep-ph",
    reportNumber = "CERN-PH-TH-2010-057, SACLAY-T10-025, IFUP-TH-2010-44",
    doi = "10.1088/1475-7516/2012/10/E01",
    journal = "JCAP",
    volume = "03",
    pages = "051",
    year = "2011",
    note = "[Erratum: JCAP 10, E01 (2012)]"
}

@article{Bringmann:2012ez,
  author = "Bringmann, Torsten and Weniger, Christoph",
  title = "{Gamma Ray Signals from Dark Matter: Concepts, Status and Prospects}",
  journal = "Phys. Dark Univ.",
  volume = "1",
  pages = "194--217",
  year = "2012",
  eprint = "1208.5481",
  archivePrefix = "arXiv",
  primaryClass = "hep-ph"
}

@article{Slatyer:2015jla,
  author = "Slatyer, Tracy R.",
  title = "{Indirect Dark Matter Signatures in the Cosmic Dark Ages. I. Generalizing the Bound on s-wave Annihilation from Planck}",
  journal = "Phys. Rev. D",
  volume = "93",
  number = "2",
  pages = "023527",
  year = "2016",
  eprint = "1506.03811",
  archivePrefix = "arXiv",
  primaryClass = "astro-ph.CO"
}

@article{Donato:2008jk,
  author = "Donato, Fiorenza and Fornengo, Nicolao and Maurin, David and Salati, Pierre",
  title = "{Antiprotons in cosmic rays from neutralino annihilation}",
  journal = "Phys. Rev. D",
  volume = "69",
  pages = "063501",
  year = "2004",
  eprint = "astro-ph/0306207"
}

@article{DiMauro:2014pqa,
    author = "Di Mauro, M. and Donato, F. and Fornengo, N. and Lineros, R. and Vittino, A.",
    title = "{Interpretation of AMS-02 electrons and positrons data}",
    eprint = "1402.0321",
    archivePrefix = "arXiv",
    primaryClass = "astro-ph.HE",
    reportNumber = "LAPTH-008-14, SACLAY-T14-010",
    doi = "10.1088/1475-7516/2014/04/006",
    journal = "JCAP",
    volume = "04",
    pages = "006",
    year = "2014"
}

@article{Boudaud:2014qra,
  author = "Boudaud, Mathieu and others",
  title = "{A new look at the cosmic ray positron fraction}",
  journal = "Astron. Astrophys.",
  volume = "575",
  pages = "A67",
  year = "2015",
  eprint = "1410.3799",
  archivePrefix = "arXiv",
  primaryClass = "astro-ph.HE"
}

@article{Abbasi:2021xzx,
    author = "Abbasi, R. and others",
    collaboration = "IceCube",
    title = "{Search for GeV-scale dark matter annihilation in the Sun with IceCube DeepCore}",
    eprint = "2111.09970",
    archivePrefix = "arXiv",
    primaryClass = "astro-ph.HE",
    doi = "10.1103/PhysRevD.105.062004",
    journal = "Phys. Rev. D",
    volume = "105",
    number = "6",
    pages = "062004",
    year = "2022"
}

@article{Adrian-Martinez:2016fei,
  author = "Adrian-Martinez, S. and others",
  collaboration = "KM3NeT",
  title = "{Letter of intent for KM3NeT 2.0}",
  journal = "J. Phys. G",
  volume = "43",
  number = "8",
  pages = "084001",
  year = "2016",
  eprint = "1601.07459",
  archivePrefix = "arXiv",
  primaryClass = "astro-ph.IM"
}

@article{Chan:2015tna,
  author = "Chan, T. K. and others",
  title = "{The impact of baryonic physics on the structure of dark matter haloes: the view from the FIRE cosmological simulations}",
  journal = "Mon. Not. Roy. Astron. Soc.",
  volume = "454",
  number = "3",
  pages = "2981--3001",
  year = "2015",
  eprint = "1507.02282",
  archivePrefix = "arXiv",
  primaryClass = "astro-ph.GA"
}

@article{Abdallah:2016ygi,
  author = "Abdallah, H. and others",
  collaboration = "H.E.S.S.",
  title = "{Search for dark matter annihilations towards the inner Galactic halo from 10 years of observations with H.E.S.S.}",
  journal = "Phys. Rev. Lett.",
  volume = "117",
  number = "11",
  pages = "111301",
  year = "2016",
  eprint = "1607.08142",
  archivePrefix = "arXiv",
  primaryClass = "astro-ph.HE"
}

@article{CTAConsortium:2018tzg,
  author = "Acharyya, A. and others",
  collaboration = "CTA Consortium",
  title = "{Sensitivity of the Cherenkov Telescope Array to a dark matter signal from the Galactic centre}",
  journal = "JCAP",
  volume = "01",
  pages = "057",
  year = "2021",
  eprint = "2007.16129",
  archivePrefix = "arXiv",
  primaryClass = "astro-ph.HE"
}

@article{Merritt:2004xa,
  author = "Merritt, David and Harfst, Stefan and Bertone, Gianfranco",
  title = "{Collisionally regenerated dark matter structures in galactic nuclei}",
  journal = "Phys. Rev. D",
  volume = "75",
  pages = "043517",
  year = "2007",
  eprint = "astro-ph/0610425"
}

@article{Amaro-Seoane:2004nt,
    author = "Amaro-Seoane, Pau and Spurzem, Rainer",
    title = "{The loss-cone problem in dense nuclei}",
    eprint = "astro-ph/0105251",
    archivePrefix = "arXiv",
    doi = "10.1046/j.1365-8711.2001.04799.x",
    journal = "Mon. Not. Roy. Astron. Soc.",
    volume = "327",
    pages = "995",
    year = "2001"
}

@article{Naoz:2019,
  author = {S. Naoz and C. M. Will and E. Ramirez-Ruiz and A. Hees and A. M. Ghez and T. Do},
  title = {A hidden friend for the Galactic Center black hole, Sgr A*},
  journal = {Astrophys. J. Lett.},
  volume = {888},
  pages = {L8},
  year = {2020},
  eprint = {1912.04910},
  archivePrefix = {arXiv},
  primaryClass = {astro-ph.GA}
}

@article{Kormendy:1995yz,
    author = "Kormendy, John and Richstone, Douglas",
    title = "{Inward bound: The Search for supermassive black holes in galactic nuclei}",
    doi = "10.1146/annurev.aa.33.090195.003053",
    journal = "Ann. Rev. Astron. Astrophys.",
    volume = "33",
    pages = "581",
    year = "1995"
}

@article{Lotz:2011av,
  author = {J. M. Lotz and P. Jonsson and T. J. Cox and D. Croton and J. R. Primack and R. S. Somerville and others},
  title = {The Major and Minor Galaxy Merger Rates at z < 1.5},
  journal = {Astrophys. J.},
  volume = {742},
  pages = {103},
  year = {2011},
  eprint = {1108.2508},
  archivePrefix = {arXiv},
  primaryClass = {astro-ph.CO}
}

@article{Reid:2004,
       author = {{Reid}, M.~J. and {Brunthaler}, A.},
        title = "{The Proper Motion of Sagittarius A*. II. The Mass of Sagittarius A*}",
      journal = {\apj},
         year = 2004,
        month = dec,
       volume = {616},
       number = {2},
        pages = {872-884},
          doi = {10.1086/424960},
archivePrefix = {arXiv},
       eprint = {astro-ph/0408107},
 primaryClass = {astro-ph},
       adsurl = {https://ui.adsabs.harvard.edu/abs/2004ApJ...616..872R}
}

@article{EHT:2022,
    author = "Akiyama, Kazunori and others",
    collaboration = "Event Horizon Telescope",
    title = "{First Sagittarius A* Event Horizon Telescope Results. V. Testing Astrophysical Models of the Galactic Center Black Hole}",
    eprint = "2311.09478",
    archivePrefix = "arXiv",
    primaryClass = "astro-ph.HE",
    reportNumber = "FERMILAB-PUB-22-419-PPD",
    doi = "10.3847/2041-8213/ac6672",
    journal = "Astrophys. J. Lett.",
    volume = "930",
    number = "2",
    pages = "L16",
    year = "2022"
}

@misc{LISA:2017,
  title = {Laser Interferometer Space Antenna (LISA) mission proposal},
  howpublished = {\url{https://www.cosmos.esa.int/web/lisa}},
  year = {2017}
}

@article{Gillessen:2017,
       author = {{Gillessen}, S. and {Plewa}, P.~M. and {Eisenhauer}, F. and {Sari}, R. and {Waisberg}, I. and {Habibi}, M. and {Pfuhl}, O. and {George}, E. and {Dexter}, J. and {von Fellenberg}, S. and {Ott}, T. and {Genzel}, R.},
        title = "{An Update on Monitoring Stellar Orbits in the Galactic Center}",
      journal = {\apj},
         year = 2017,
        month = mar,
       volume = {837},
       number = {1},
          eid = {30},
        pages = {30},
          doi = {10.3847/1538-4357/aa5c41},
archivePrefix = {arXiv},
       eprint = {1611.09144},
 primaryClass = {astro-ph.GA},
       adsurl = {https://ui.adsabs.harvard.edu/abs/2017ApJ...837...30G}
}

@article{Gravity:2020,
    author = "Abuter, R. and others",
    collaboration = "GRAVITY",
    title = "{Detection of the Schwarzschild precession in the orbit of the star S2 near the Galactic centre massive black hole}",
    eprint = "2004.07187",
    archivePrefix = "arXiv",
    primaryClass = "astro-ph.GA",
    doi = "10.1051/0004-6361/202037813",
    journal = "Astron. Astrophys.",
    volume = "636",
    pages = "L5",
    year = "2020"
}

@book{BinneyTremaine,
       author = {{Binney}, James and {Tremaine}, Scott},
        title = "{Galactic Dynamics: Second Edition}",
         year = 2008,
       adsurl = {https://ui.adsabs.harvard.edu/abs/2008gady.book.....B}
}

@article{Merritt:2010,
    author = "Merritt, David",
    title = "{Evolution of Nuclear Star Clusters}",
    eprint = "0802.3186",
    archivePrefix = "arXiv",
    primaryClass = "astro-ph",
    doi = "10.1088/0004-637X/694/2/959",
    journal = "Astrophys. J.",
    volume = "694",
    pages = "959--970",
    year = "2009"
}

@article{Antonini:2012ad,
  author = {Antonini, F. and Merritt, D.},
  title = {Dynamical Friction around Supermassive Black Holes},
  journal = {Astrophys. J.},
  volume = {745},
  pages = {83},
  year = {2012},
  eprint = {1108.1163},
  archivePrefix = {arXiv},
  primaryClass = {astro-ph.GA}
}

@article{Gualandris:2010,
       author = {{Gualandris}, Alessia and {Merritt}, David},
        title = "{Long-term Evolution of Massive Black Hole Binaries. IV. Mergers of Galaxies with Collisionally Relaxed Nuclei}",
      journal = {\apj},
         year = 2012,
        month = jan,
       volume = {744},
       number = {1},
          eid = {74},
        pages = {74},
          doi = {10.1088/0004-637X/744/1/74},
archivePrefix = {arXiv},
       eprint = {1107.4095},
 primaryClass = {astro-ph.GA},
       adsurl = {https://ui.adsabs.harvard.edu/abs/2012ApJ...744...74G}
}

@article{Profumo:2025prep,
  author = {S. Profumo and A. Liu and P. Giffin and others},
  title = {The Dark Companion and the Scouring of the Galactic Center Dark Matter Spike},
  journal = {in preparation},
  year = {2025}
}

@article{Shapiro:2016ypb,
  author = {Shapiro, Stuart L. and Shelton, Jessie},
  title = {Weak annihilation cusp inside the dark matter spike about a black hole},
  journal = {Phys. Rev. D},
  volume = {93},
  pages = {123510},
  year = {2016},
  eprint = {1606.01248},
  archivePrefix = {arXiv},
  primaryClass = {astro-ph.HE}
}

@article{Merritt:2006mt,
  author = {Merritt, David},
  title = {Mass deficits, stalling radii, and the merger histories of elliptical galaxies},
  journal = {Astrophys. J.},
  volume = {648},
  pages = {976--986},
  year = {2006},
  eprint = {astro-ph/0603439}
}

@article{Thomas:2014,
    author = "Thomas, J. and Saglia, R. P. and Bender, R. and Erwin, P. and Fabricius, M.",
    title = "{The Dynamical Fingerprint of Core Scouring in Massive Elliptical Galaxies}",
    eprint = "1311.3783",
    archivePrefix = "arXiv",
    primaryClass = "astro-ph.GA",
    doi = "10.1088/0004-637X/782/1/39",
    journal = "Astrophys. J.",
    volume = "782",
    number = "1",
    pages = "39",
    year = "2014"
}

@article{Bortolas:2016,
    author = "Bortolas, E. and Gualandris, A. and Dotti, M. and Spera, M. and Mapelli, M.",
    title = "{Brownian motion of massive black hole binaries and the final parsec problem}",
    eprint = "1606.06728",
    archivePrefix = "arXiv",
    primaryClass = "astro-ph.GA",
    doi = "10.1093/mnras/stw1372",
    journal = "Mon. Not. Roy. Astron. Soc.",
    volume = "461",
    number = "1",
    pages = "1023--1031",
    year = "2016"
}

@article{Roedig:2011,
  author       = {Roedig, C. and Dotti, M. and Sesana, A. and Cuadra, J. and Colpi, M.},
  title        = {Limiting eccentricity of subparsec massive black hole binaries surrounded by self-gravitating gas discs},
  journal      = {Mon. Not. Roy. Astron. Soc.},
  volume       = {415},
  pages        = {3033--3041},
  year         = {2011},
  eprint       = {1104.3868},
  archivePrefix = {arXiv},
  primaryClass = {astro-ph.CO},
  doi          = {10.1111/j.1365-2966.2011.18927.x}
}

@article{Unavane:1996,
  author       = {Unavane, M. and Wyse, R. F. G. and Gilmore, G.},
  title        = {Galactic halo formation in the context of hierarchical galaxy formation models},
  journal      = {Mon. Not. Roy. Astron. Soc.},
  volume       = {278},
  pages        = {727--736},
  year         = {1996},
  doi          = {10.1093/mnras/278.3.727}
}

@article{Ghez:1998,
  author       = {Ghez, A. M. and Klein, B. L. and Morris, M. and Becklin, E. E.},
  title        = {High proper-motion stars in the vicinity of Sagittarius A*: Evidence for a supermassive black hole at the center of our Galaxy},
  journal      = {Astrophys. J.},
  volume       = {509},
  pages        = {678--686},
  year         = {1998},
  eprint       = {astro-ph/9807210},
  doi          = {10.1086/306528}
}

@article{Feldmeier-Krause:2017,
       author = {{Feldmeier-Krause}, A. and {Zhu}, L. and {Neumayer}, N. and {van de Ven}, G. and {de Zeeuw}, P.~T. and {Sch{\"o}del}, R.},
        title = "{Triaxial orbit-based modelling of the Milky Way Nuclear Star Cluster}",
      journal = {\mnras},
         year = 2017,
        month = apr,
       volume = {466},
       number = {4},
        pages = {4040-4052},
          doi = {10.1093/mnras/stw3377},
archivePrefix = {arXiv},
       eprint = {1701.01583},
 primaryClass = {astro-ph.GA},
       adsurl = {https://ui.adsabs.harvard.edu/abs/2017MNRAS.466.4040F}
}

@article{Schodel:2014,
  author = {Schödel, R. and Feldmeier, A. and Kunneriath, D. and Stolovy, S. and Neumayer, N. and Amaro-Seoane, P. and Nishiyama, S.},
  title = {The nuclear cluster of the Milky Way: Our primary testbed for the interaction of a dense star cluster with a massive black hole},
  journal = {Class. Quant. Grav.},
  volume = {31},
  pages = {244007},
  year = {2014},
  eprint = {1411.4504},
  archivePrefix = {arXiv},
  primaryClass = {astro-ph.GA}
}

@article{Ciambur:2017,
       author = {{Ciambur}, Bogdan C. and {Graham}, Alister W. and {Bland-Hawthorn}, Joss},
        title = "{Quantifying the (X/peanut)-shaped structure of the Milky Way - new constraints on the bar geometry}",
      journal = {\mnras},
         year = 2017,
        month = nov,
       volume = {471},
       number = {4},
        pages = {3988-4004},
          doi = {10.1093/mnras/stx1823},
archivePrefix = {arXiv},
       eprint = {1706.09902},
 primaryClass = {astro-ph.GA},
       adsurl = {https://ui.adsabs.harvard.edu/abs/2017MNRAS.471.3988C}
}

@article{Schodel:2009,
       author = {{Sch{\"o}del}, R. and {Merritt}, D. and {Eckart}, A.},
        title = "{The nuclear star cluster of the Milky Way: proper motions and mass}",
      journal = {\aap},
         year = 2009,
        month = jul,
       volume = {502},
       number = {1},
        pages = {91-111},
          doi = {10.1051/0004-6361/200810922},
archivePrefix = {arXiv},
       eprint = {0902.3892},
 primaryClass = {astro-ph.GA},
       adsurl = {https://ui.adsabs.harvard.edu/abs/2009A&A...502...91S}
}

@article{Gallego-Cano:2018,
       author = {{Gallego-Cano}, E. and {Sch{\"o}del}, R. and {Dong}, H. and {Nogueras-Lara}, F. and {Gallego-Calvente}, A.~T. and {Amaro-Seoane}, P. and {Baumgardt}, H.},
        title = "{The distribution of stars around the Milky Way's central black hole. I. Deep star counts}",
      journal = {\aap},
         year = 2018,
        month = jan,
       volume = {609},
          eid = {A26},
        pages = {A26},
          doi = {10.1051/0004-6361/201730451},
archivePrefix = {arXiv},
       eprint = {1701.03816},
 primaryClass = {astro-ph.GA},
       adsurl = {https://ui.adsabs.harvard.edu/abs/2018A&A...609A..26G}
}

@article{Sofue:2013,
       author = {{Sofue}, Yoshiaki},
        title = "{Rotation Curve and Mass Distribution in the Galactic Center -From Black Hole to Entire Galaxy-}",
      journal = {\pasj},
         year = 2013,
        month = dec,
       volume = {65},
       number = {6},
          eid = {118},
        pages = {118},
          doi = {10.1093/pasj/65.6.118},
archivePrefix = {arXiv},
       eprint = {1307.8241},
 primaryClass = {astro-ph.GA},
       adsurl = {https://ui.adsabs.harvard.edu/abs/2013PASJ...65..118S}
}

@article{Launhardt:2002,
    author = "Launhardt, R. and Zylka, R. and Mezger, P. G.",
    title = "{The nuclear bulge of the galaxy. 3. Large scale physical characteristics of stars and interstellar matter}",
    eprint = "astro-ph/0201294",
    archivePrefix = "arXiv",
    doi = "10.1051/0004-6361:20020017",
    journal = "Astron. Astrophys.",
    volume = "384",
    pages = "112--139",
    year = "2002"
}

@article{Wyrzykowski:2015,
       author = {{Wyrzykowski}, L. and {Skowron}, J. and {Koz{\l}owski}, S. and {Udalski}, A. and {Szyma{\'n}ski}, M.~K. and {Kubiak}, M. and {Pietrzy{\'n}ski}, G. and {Soszy{\'n}ski}, I. and {Szewczyk}, O. and {Ulaczyk}, K. and {Poleski}, R. and {Tisserand}, P.},
        title = "{The OGLE view of microlensing towards the Magellanic Clouds - IV. OGLE-III SMC data and final conclusions on MACHOs}",
      journal = {\mnras},
         year = 2011,
        month = oct,
       volume = {416},
       number = {4},
        pages = {2949-2961},
          doi = {10.1111/j.1365-2966.2011.19243.x},
archivePrefix = {arXiv},
       eprint = {1106.2925},
 primaryClass = {astro-ph.GA},
       adsurl = {https://ui.adsabs.harvard.edu/abs/2011MNRAS.416.2949W}
}

@article{Bennett:2019,
       author = {{Sumi}, T. and {Bennett}, D.~P. and {Bond}, I.~A. and {Abe}, F. and {Botzler}, C.~S. and {Fukui}, A. and {Furusawa}, K. and {Itow}, Y. and {Ling}, C.~H. and {Masuda}, K. and {Matsubara}, Y. and {Muraki}, Y. and {Ohnishi}, K. and {Rattenbury}, N. and {Saito}, To. and {Sullivan}, D.~J. and {Suzuki}, D. and {Sweatman}, W.~L. and {Tristram}, P.~J. and {Wada}, K. and {Yock}, P.~C.~M. and {MOA Collaboratoin}, The},
        title = "{The Microlensing Event Rate and Optical Depth toward the Galactic Bulge from MOA-II}",
      journal = {\apj},
         year = 2013,
        month = dec,
       volume = {778},
       number = {2},
          eid = {150},
        pages = {150},
          doi = {10.1088/0004-637X/778/2/150},
archivePrefix = {arXiv},
       eprint = {1305.0186},
 primaryClass = {astro-ph.GA},
       adsurl = {https://ui.adsabs.harvard.edu/abs/2013ApJ...778..150S}
}

@article{Portail:2017,
  author = {Portail, M. and Wegg, C. and Gerhard, O. and Ness, M.},
  title = {Dynamical modelling of the Galactic bulge and bar: The pattern speed, stellar, and dark matter mass distribution},
  journal = {Mon. Not. Roy. Astron. Soc.},
  volume = {465},
  pages = {1621--1644},
  year = {2017},
  eprint = {1608.07954},
  archivePrefix = {arXiv},
  primaryClass = {astro-ph.GA}
}

@article{El-Zant:2001,
  author = {El-Zant, A. and Shlosman, I. and Hoffman, Y.},
  title = {Dark halos: The flattening of the density cusp by dynamical friction},
  journal = {Astrophys. J.},
  volume = {560},
  pages = {636--643},
  year = {2001},
  eprint = {astro-ph/0103386}
}

@article{Lee:2015fea,
  author = {Lee, S. K. and Lisanti, M. and Safdi, B. R. and Slatyer, T. R. and Xue, W.},
  title = {Evidence for Unresolved Gamma-Ray Point Sources in the Inner Galaxy},
  journal = {Phys. Rev. Lett.},
  volume = {116},
  number = {5},
  pages = {051103},
  year = {2016},
  eprint = {1506.05124},
  archivePrefix = {arXiv},
  primaryClass = {astro-ph.HE}
}

@article{Bartels:2018,
    author = "Bartels, Richard and Storm, Emma and Weniger, Christoph and Calore, Francesca",
    title = "{The Fermi-LAT GeV excess as a tracer of stellar mass in the Galactic bulge}",
    eprint = "1711.04778",
    archivePrefix = "arXiv",
    primaryClass = "astro-ph.HE",
    doi = "10.1038/s41550-018-0531-z",
    journal = "Nature Astron.",
    volume = "2",
    number = "10",
    pages = "819--828",
    year = "2018"
}

@article{Ahnen:2017,
    author = "Ahnen, M. L. and others",
    collaboration = "MAGIC, also at Japanese MAGIC Consortium",
    title = "{Search for very high-energy gamma-ray emission from the microquasar Cygnus X-1 with the MAGIC telescopes}",
    eprint = "1708.03689",
    archivePrefix = "arXiv",
    primaryClass = "astro-ph.HE",
    doi = "10.1093/mnras/stx2087",
    journal = "Mon. Not. Roy. Astron. Soc.",
    volume = "472",
    number = "3",
    pages = "3474--3485",
    year = "2017"
}

@inproceedings{IceCubeGen2:2023,
    author = "Ishihara, Aya and others",
    collaboration = "IceCube-Gen2",
    title = "{The next generation neutrino telescope: IceCube-Gen2}",
    eprint = "2308.09427",
    archivePrefix = "arXiv",
    primaryClass = "astro-ph.HE",
    reportNumber = "PoS-ICRC2023-994",
    doi = "10.22323/1.444.0994",
    journal = "PoS",
    volume = "ICRC2023",
    pages = "994",
    year = "2023"
}

@article{KM3NeT:2016LoI,
  author         = {Adri{\'a}n-Mart{\'i}nez, S. and others},
  collaboration  = {KM3NeT},
  title          = {Letter of Intent for {KM3NeT} 2.0},
  journal        = {J.\ Phys.\ G},
  volume         = {43},
  number         = {8},
  pages          = {084001},
  year           = {2016},
  doi            = {10.1088/0954-3899/43/8/084001},
  eprint         = {1601.07459},
  archivePrefix  = {arXiv},
  primaryClass   = {astro-ph.IM}
}

@article{SWGO:2019WhitePaper,
    author = "Abreu, P. and others",
    title = "{The Southern Wide-Field Gamma-Ray Observatory (SWGO): A Next-Generation Ground-Based Survey Instrument for VHE Gamma-Ray Astronomy}",
    eprint = "1907.07737",
    archivePrefix = "arXiv",
    primaryClass = "astro-ph.IM",
    month = "7",
    journal = "",
    year = "2019"
}

@article{NBodySpike2025,
    author = "Dosopoulou, Fani and Silk, Joseph",
    title = "{Multimessenger Detection of Black Hole Binaries in Dark Matter Spikes}",
    eprint = "2502.15468",
    archivePrefix = "arXiv",
    primaryClass = "astro-ph.HE",
    doi = "10.1103/hysg-5273",
    journal = "Phys. Rev. Lett.",
    volume = "135",
    number = "8",
    pages = "081401",
    year = "2025"
}

\end{document}